# Diagnostics of collisions between electrons and water molecules in near-ultraviolet and visible wavelengths

Short Title: Diagnostics of electron impact on water molecules

D. Bodewits[1], J. Országh[2], J. Noonan[3], M. Ďurian[2], Š. Matejčík[2]

[1]Physics Department, Leach Science Center, Auburn University, Auburn, AL 36832, USA; dennis@auburn.edu

[2]Department of Experimental Physics, Faculty of Mathematics, Physics and Informatics, Comenius University in Bratislava, Mlynská dolina F-2, 842 48 Bratislava; Slovak Republic; juraj.orszagh@uniba.sk, matejcik@fmph.uniba.sk

[3]Lunar and Planetary Laboratory, University of Arizona, 1629 E University Boulevard, Tucson, AZ 85721-0092, USA

**Abstract**

We studied dissociation reactions of electron impact on water vapor for several fragment species at optical and near ultraviolet wavelengths (200 – 850 nm). The resulting spectrum is dominated by the Hydrogen Balmer series, by the OH (A $^2\Sigma^+$ – X $^2\Pi$) band, and by the emission of ionic $H_2O^+$ (A $^2A_1$ – X $^2B_1$) and $OH^+$ (A $^3\Pi$ – X $^3\Sigma^-$) band systems. Emission cross sections and reaction channel thresholds were determined for energies between 5 – 100 eV. We find that electron impact dissociation of $H_2O$ results in an emission spectrum of the OH (A $^2\Sigma^+$ – X $^2\Pi$) band that is distinctly different than the emission spectra from other excitation mechanisms seen in planetary astronomy. We attribute the change to a strongly non-thermal population of rotational states seen in planetary astronomy. This difference can be utilized for remote probing of the contribution of different physical reactions in astrophysical environments.

## 1. INTRODUCTION

Dissociative electron impact excitation reactions can provide a remote diagnostic of neutral gases and the physical environment of atmospheres around planets and small bodies in our solar system (Galand et al. 2002). It provides distinct spectral fingerprints in the infrared, optical, and ultraviolet wavelengths, and its efficiency is strongly dependent on both the electrons' energy and on the target molecule (Itikawa & Mason 2005). As such, electron dissociative excitation and subsequent emission has been used to identify water plumes emanating from Europa (Hall et al. 1995; Roth et al. 2014), to remotely identify a tenuous $O_2$ atmosphere around Callisto (Cunningham et al. 2015), and recently, by the *Rosetta* orbiter to study neutral gases in the inner coma of 67P/Churyumov-Gerasimenko (Feldman et al. 2015, 2018; Bodewits et al. 2016) and its interaction with a coronal mass ejection (Noonan et al. 2018).

Both the Europa and 67P/Churyumov-Gerasimenko results have sparked new interest in electron impact reactions with molecules abundant in the atmospheres of small bodies (including $H_2O$, $CO_2$, CO, HCN, $O_2$, …) at energies between 0 – 100 eV. They also highlighted gaps and limitations in existing data sets (Van de Burgt et al. 1989; Avakyan et al. 1998; Itikawa & Mason 2005; McConkey et al. 2008). Cross sections for several excitation-emission features are often measured only at one or two electron energies, many times only relative cross sections are available, or cross sections vary greatly between different groups.

There are large differences between measurements caused by the different methods used (for example crossed beam vs. gas cell experiments), overlapping emission features that make it difficult to measure the entire emission cross section of a given excited product, ambiguities in data processing (correction for lifetimes), the effects of energy resolution and the calibration of the electron sources used, and even the temperature of the target gas. For example, Table 6 in Van de Burgt et al. 1989 shows a factor of four difference across measurements of the H I, O I 121.7 nm optical excitation function for electrons colliding with $H_2O$, even after renormalization. A similar issue can also be seen with O I 130.4 nm in both Van de Burgt et al. (1989) and Makarov et al. (2004). To make full use of the diagnostic qualities of electron impact emission, reliable measurements of the energy dependence of emission features in both visible and ultraviolet wavelengths are needed, at near-threshold electron energies.

Here, we report on laboratory measurements of electron impact reactions with water vapor at energies relevant for the atmospheres of small bodies such as comets (Broiles et al. 2016) and Europa (75% of electrons below 205 eV; Bagenal et al. 2014). We first briefly describe the experimental apparatus and our data reduction (Sec. 2), then we discuss the measured cross sections and reaction channel thresholds (Sec. 3). We conclude this manuscript with a discussion of the diagnostic application to small body (cometary) atmospheres (Sec. 4).

## 2. EXPERIMENT AND ANALYSIS

*2.1 Electron fluorescence apparatus*

The experiments were carried out using a crossed-beam apparatus using distilled and deionized water. The apparatus is described in detail in previous publications (Danko et al. 2013; Országh et al. 2017) and is briefly described here. The electron beam generated by an electron gun crosses perpendicularly with a molecular beam formed by an effusive capillary in the vacuum chamber. In the experiments discussed in this manuscript we used the trochoidal electron monochromator (TEM) only for a small number of measurements due to the low intensity of produced electron beam and low values of the emission cross sections of the reactions leading to weak photon signal. The entire chamber was heated to approximately 60°C to prevent condensation of $H_2O$ vapor. We ensured that gas pressures did not exceed the single collision regime - an electron hits only one molecule. The energy resolution of the electron beam produced with the TEM was 600 meV FWHM with electron currents typically between 0.3 – 2 µA. The electron gun had an electron energy distribution with a FWHM of 3 eV and currents between 5 – 8 µA. For both sources, the electron energy range considered was 5 – 100 eV. The gas pressure, ambient pressure, and electron current were all electronically monitored and logged.

Photons produced by the reactions were guided out of the vacuum chamber by a system of mirrors and lenses and focused onto the entrance slit of the Czerny-Turner optical monochromator (resolution $\lambda/d\lambda$ = 972 at 100 μm slit width). To acquire a broad-range overview spectrum we used a Hamamatsu R3896 photomultiplier tube sensitive between 185 – 900 nm. To acquire more detailed spectra, we used a Hamamatsu R4220P photomultiplier tube which is more sensitive and has lower noise in the UV range than the R3896 photomultiplier. We determined the spectral response function of the optical system in the visible and near infrared by measuring black body radiation emitted by a heated tungsten filament of known temperature. We determined the spectral sensitivity response in the ultraviolet range by measuring the continuum emission of $H_2$ (a $^3\Sigma_g^+$ − b $^3\Sigma_u^+$) at 14 eV. The shape of the hydrogen continuum spectrum published by James et al. 1998 and the theoretical spectrum of the black body were used as a reference each in its corresponding spectral range. The final spectral response function was determined as a ratio of measured spectra and reference spectra. The instrumental field of view is given by the optical monochromator acceptance angle and the parameters and positions of the used lenses and mirror. The interaction region from which the emitted photons can reach the photomultiplier is approximately 3 mm in diameter for a 100 μm slit width.

Two complementary methods were applied to measure electron collision energy dependent emission spectra: 1. wavelength scans at a fixed electron energy, and 2. electron energy scans at a fixed wavelength corresponding to specific emission features of interest. In both cases, typical exposure times were 10 seconds per measurement step. Profiles of the emission intensity with respect to electron incident energy were derived by averaging over at least three scans.

## 2.2 Calibration Procedures

To ensure good signal to noise for the optical monochromator we used relatively large background gas pressures in the reaction chamber (~$10^{-4}$ mbar). At the pressures used for our experiments polar water molecules form a deposit on the electrodes of the electron gun (Berman 1996) which causes a linear offset in the electron energy. To correct for this the energy of the electron beam was first absolutely calibrated by introducing a mixture of $N_2$, helium, and $H_2O$ into the set up and by measuring the intensity profile of the $N_2$ ($C^3\Pi_u$–$B^3\Pi_g$)(0–0) band at 337 nm and the He I (1s2p $^3P_{1,2}^0$− 1s4d $^3D_{1,2,3}$−) emission line at 447.14 nm. The former has a sharp maximum at an electron energy of 14.1 eV (Országh et al. 2012), the latter has a threshold at an electron energy of 23.736 eV (NIST 2015). We then measured the energy dependence of the intensity of the $H_\beta$ emission for both the gas mixture and the pure $H_2O$ beam. By fitting the $H_2O$ -only $H_\beta$ profile to the $H_\beta$ profile of the $H_2O$ /$N_2$/He mixture, we corrected for the linear offset of the electron energy in our $H_2O$ experiments.

The energy dependence of the emission lines intensities was determined by positioning the optical monochromator at the peak of the lines, and then scanning over the electron energy range. The absolute calibration of the excitation curve was achieved by scaling the curve of the $H_\beta$ line at 100 eV to the measurements by Müller et al. (1993), 4.9 x $10^{-19}$ cm². Then the intensity of the spectrum was scaled such that the area below the spectral line corresponded to the value of the cross section curve at 50 eV. In this procedure we assume that only the relative cross section and not the shape of the atomic line changes with electron energy. Other emission lines were calibrated by scaling the data according to $H_\beta$.

To determine the emission cross section of the much broader molecular emission features, the area under the band was integrated after the spectral intensity was calibrated according to $H_\beta$. Specific emission cross section normalization details are discussed later in paragraphs corresponding to the individual cross sections.

### 2.3 Thresholds

Several steps are visible in the relation between emission cross sections and the electron energy. These steps or onsets indicate the thresholds for opening of new reaction channels for given dissociative excitation processes. We determined the position of these threshold energies by fitting a theoretical threshold function consisting of a constant background and a linear function. The intersection of the two functions determines the threshold. More sophisticated routines are available, but they require data with higher signal-to-noise ratio. The noise in the measured signal affects the uncertainty of the determined value more significantly than the uncertainty introduced by the fitting method.

### 2.4 Uncertainties

The measured emission cross sections are subject to several uncertainties. The largest contribution to the uncertainty in the cross sections originates from the spectral intensity calibration process in which we anchor our data to the dissociative emission cross section of the H$_\beta$ line produced by electron impact on water vapor by Muller et al. (1993), who claimed a systematic uncertainty of only 10% (see Sec. 2.2). As discussed in Sec. 1, results reported by other groups vary widely. This uncertainty affects all data points and leads to a simple scaling factor.

The uncertainty of the sensitivity of the spectrometer is between 20% for the shortest (200 – 250 nm) and longest (750 – 850 nm) wavelengths and is approximately 5% for the middle section (250 – 750 nm) of the region.

The data were not corrected for electron beam and target pressure variations, which both introduce a random error of approximately 5%. Both the pressure and electron current were logged and measurements where either of these varied by more than 10% were discarded. Every scan over electron energy was repeated at least 5 times and the standard deviation in signal counts was used to determine that the stochastical errors in every energy bin was on the order of 5%.

No method for determination of polarization of the emitted light was used in this experiment, introducing an error of less than 5%.

The long lifetime of the higher $n \geq 5$ states of the hydrogen atom results in a large fraction of the Balmer emission series not being detected (Sec. 2.2). We corrected our measurements for this assuming a velocity of 7 km/s; in reality the fragments' velocity distributions also have faster components that are added with increasing collision energy (Kouchi et al. 1979; Kurawaki et al. 1993; Makarov et al. 2004). Consequently, the cross sections of the H$_{\gamma,\delta,\varepsilon}$ may be underestimated somewhat at higher collision energies.

Finally, the uncertainty of the electron energy calibration is mostly driven by the energy distribution of the electron beam which has a full width half maximum (FWHM) of 0.6 eV, equivalent to a 1-sigma error of 0.26 eV.

## 3. RESULTS AND DISCUSSION

We measured the electron induced emission spectrum of water at an incident electron energy of 50 eV in the spectral region from 200 to 800 nm. An overview spectrum (300 – 800 nm) is presented in Fig. 1. It was acquired at a lower spectral resolution of 1 nm with wider optical monochromator slits of 300 um. The main features of the spectrum are the OH (A $^2\Sigma^+$ – X $^2\Pi$) bands and the Hydrogen Balmer lines produced by dissociative electron impact of water molecules. In addition, weaker bands produced by electron dissociation and ionization processes were detected, namely OH$^+$ (A $^3\Pi$ – X $^3\Sigma^-$) and H$_2$O$^+$ (A $^2A_1$ – X $^2B_1$). Finally, we marginally detected an emission line from atomic oxygen at 777 nm (where our detector has lower sensitivity, corresponding to the OI (3s$^5$S$^0$ – 3p$^5$P) transition. The [OI] emission lines from the forbidden transitions at 557.7, 630.0, and 636.4 nm are not detected in our experiment, owing to the long lifetimes of the metastable (2P$^4$)$^1$S and (2P$^4$)$^1$D states (cf. Bhardwaj & Rahuram 2012).

### 3.1 Hydrogen Balmer series

In our spectra we can reliably detect the Balmer series up to H$_\varepsilon$. The higher order transitions of the Balmer series have a relative long lifetime (Table 1) allowing a significant fraction of the hydrogen atoms to leave the field of view (diameter 3 mm) before emitting a photon. Doppler profile measurements (not possible with our spectral resolution) and translational energy spectroscopy measurements indicate that the hydrogen atoms have significant velocities throughout the range of 6 – 8 km/s (0.19 – 0.34 eV), and that their velocity distribution changes when additional reaction channels open (Kouchi et al. 1979; Kurawaki et al. 1983; Makarov et al. 2004). This implies that all the H$_\alpha$ decays within our field of view, but that as much as 91% of the H$_\varepsilon$ emission is not detected. In published studies, it is often unclear what correction is applied, if any at all.

The spectral region between 430 and 750 nm contains many H$_2$O$^+$ sub bands (Section 3.4). The possible overlap with the H$_2$O$^+$ emission can affect the Balmer series cross sections determination. However, because of the low intensity of the H$_2$O$^+$ bands and because the much brighter hydrogen lines are very narrow the impact of the H$_2$O$^+$ background on the Balmer series is negligible.

Emission cross sections and thresholds of the Balmer series are shown in Figures 2 and 3. All show a steep increase starting around 18 eV, with the threshold increasing from 17.7 to 19.5 eV going from H$_\alpha$ to H$_\delta$. The signal-to-noise ratio for the H$_\varepsilon$ line was too low to reliably determine a first activation threshold. The thresholds are consistent with those reported in previous studies (Table 3) and are all about 0.5 eV higher than the thermochemical minimum thresholds for the reaction $H_2O + e \rightarrow OH (X\ ^2\Pi^+) + H(np)$ (Table 4). All Balmer emission cross sections show a second onset between 23.6 – 25.5 eV. This energy range is consistent with several reactions that result in complete dissociation of the water molecule and with the production of excited hydrogen and excited neutral or ionized hydroxyl. The emission cross sections profiles of the OH (A $^2\Sigma^+$ – X $^2\Pi^+$) (Sec. 3.2) and OH$^+$ (A $^2\Sigma^+$ – X $^2\Pi^+$) bands have thresholds around 9.4 and 13 eV, consistent with the production of hydrogen in the ground state. They also show small kinks at 20 and 23 eV, above the thresholds for the production of excited hydrogen occurs according to the thermochemical estimates.

Our cross section values for the Balmer series of hydrogen are in agreement with those published by Müller et al. (1993) within experimental uncertainty. The values reported by Beenakker et al. (1974) and Möhlmann et al. (1979) are in general higher and finally the values reported by Vroom et al. (1969) are significantly higher. Our experiment is based on crossed beams method, Müller et al. (1993) utilizes both crossed beams and collision cell method and in the remaining three publications the collision cell filled with gas to a specific pressure was used. The cross section values in collision cell experiments are calculated based on the measured intensity and concentration of particles in the cell. In case of Vroom et al. (1969) the concentration determination was based on measurement of the sample cylinder temperature which could negatively influence the uncertainty of cross section evaluation. Hence, the discrepancy between the cross section values published by Vroom et al. (1969) and later papers. The long lifetime of higher excited states contributing to Balmer series (given in Table 1) significantly affects the photon signal intensity in crossed beams experiment since a considerable part of the photons can be lost due to limited field of view of the experimental device (see Table 1). After application of calculated correction factors our values are comparable to Müller et al. (1993) and have similar trend as cross sections determined in collision cell experiments which supports the correctness of the used correction.

As is shown in Fig. 3A, all the curves of the emission cross sections of Balmer series show similar trends with respect to the collision energy. That agrees with Beenakker et al. (1974) who reports similarity within 4% and deduces the other Balmer series cross sections from the H (4 – 2) value. In our experiment every emission cross section of H ($n \rightarrow 2$) has been determined individually from the experimental data and the determined cross sections are in good agreement to Beenakker's values. The similar shape of the curves can be also seen from the fig. 3B where the ratios of $H_\beta/H_\alpha$ and $H_\gamma/H_\alpha$ are roughly constant after they reach a maximum at approximately 40 eV.

In general, the authors of the published data do not describe experimental details such as the temperature of water vapor reacting with electron beam that may contribute to discrepancies of the published cross section values.

### *3.2 Emission from OH*

The emission spectrum of the OH (A $^2\Sigma^+$ – X $^2\Pi$) transitions from 260 nm to 335 nm is shown in Fig. 4. This region was measured with higher spectral resolution than used for the overview spectrum in Fig. 1 and the signal accumulation time was increased to gain better signal to noise ratio. Most of the emission occurs between 305 nm and 335 nm, which contains the (0 – 0), (1 – 1), (2 – 2) and (3 – 3) bands; the section between 280 nm and 295 nm contains the (1 – 0) and (2 – 1) bands; and the faintest part of the spectrum occurs between 260 nm and 270 nm which contains the (2 – 0) band. Emission from the much fainter ionic OH$^+$ (A $^3\Pi$ – X $^3\Sigma^-$) band overlaps partially with the neutral OH (A $^2\Sigma^+$ – X $^2\Pi$) emission between 250 to 350 nm and is shown separately in Figure 4.

Our absolute calibration of the emission cross section of electron impact produced OH (A $^2\Sigma^+$ – X $^2\Pi$) is done by anchoring the flux in a small bandpass $\Delta\lambda$ to the emission cross section of $H_\beta$ and by weighting this flux to the entire emission band based on an empirical ro-vibrational model. The cross section of $H_\beta$ in turn was calibrated by comparing our signal at electron impact energies of 100 eV to the emission cross section reported by Müller et al (1993) at the same energy.

As discussed in Section 2.1, we acquire our measurements by either fixing the electron energy and by adjusting the spectrometer to acquire a spectrum over a range of wavelengths, or by fixing the spectrometer at specific wavelengths (283.5 nm and 307.3 nm in our case; Fig. 5a) and by scanning over electron impact energy. To determine the total emission cross section of the different OH bands from the fixed-wavelength measurements we employed a technique inspired by Müller et al. (1993), who fitted different simulated ro-vibrational distributions to the experimental OH spectrum to deconvolute the emission spectrum. This distribution cannot be directly determined from the experimental data due to limited spectral resolution.

To construct synthetic spectra of the sections from 280 nm to 305 nm and from 305 nm to 335 nm, we simulated the spectrum using the *Lifbase* 2.1 software (Luque & Crosley, 1999) by varying the vibrational and rotational populations for $v' = 0, 1, 2$ (Fig. 6). To match the experimental and simulated spectra it was necessary to slightly adjust the wavelength calibration of the experimental spectrum by 0.3 nm which is within the experimental resolution of the optical monochromator. The baseline correction of the experimental spectrum was set to 0.55%, a gaussian line shape was found to better fit the experimental spectrum, and the resolution of synthetic spectrum was set to 0.5 nm. Since the *Lifbase* software allows only manual adjustment of the rotational level populations in the model we have started our approximation by first setting the populations according to Müller et al. (1993). We adjusted the populations manually to achieve a minimal residual after subtracting the fitted spectra from the measurement.

In Fig. 6 we show the best-fit synthetic spectrum with the contributions of the $v'=0$, 1, and 2 bands, compared to a spectrum measured at an incident energy of 50 eV. Müller et al. (1993) also considered the $OH^+$ (A $^3\Pi$ – X $^3\Sigma^-$)(2 – 0) feature in their fit. Because the contribution of this band to the total emission is marginal and does not affect the fit to our data, we excluded it from the spectral model. In Figure 6 we show the relative population of the rotational states for individual vibrational states of OH (A $^2\Sigma^+$), in agreement with previous studies (Müller et al. 1993; Möhlmann 1976). Based on the relative strenghts of the $Q_1$ and $R_1$ bands the rotational temperature of the synthetic spectrum corresponds to approximately 2200 K.

Next, we used the synthetic spectra to extrapolate total emission cross sections from the measurements at 283.6 nm and 307.3 nm by integrating the area under the curve for the specific band. For this we assumed that the relative distribution of the rovibrational states was constant over the energy range considered here. We verified this assumption by comparing our distribution measured at 50 eV by the vibrational distribution of Müller et al. (1993) that was measured at 100 eV. Our results are very similar, except for the population of the rotational states for vibrational level $v\,'=2$. We also looked at the ratio between the measured emission cross sections at 283.6 and 307.3 nm as a function of the incident electron energy. This ratio is approximately constant within 10% supporting the validity of our assumption.

The resulting emission cross sections for the different OH (A $^2\Sigma^+$ – X $^2\Pi$) transitions at an impact energy of 50 eV are summarized in Table 5. For comparison, we also give the values determined by Müller et al. (1993) acquired at an incident electron energy 100 eV. The first three rows in the Table ($v\,'$- $v\,''$ = 0–0, 1–1, 2–2) add to the emission between 307 – 330 nm; the next two rows ($v\,'$- $v\,''$ = 1–0, 2–1) constitute the emission between 280 – 295 nm and finally the last row ($v\,'$- $v\,''$ = 2–0) corresponds to band between 260 – 270 nm (see Fig. 6). As the three individual bands that constitute each of the two emission features cannot be distinguished in the measured spectra, we add the emission cross sections of these bands to derive the 'unresolved emission cross section'. The cross sections given by Müller et al. (1993) are approximately two times larger than our values for both unresolved band and resolved bands. That corresponds to the increase of relative cross section curve at 100 eV in comparison to 50 eV even though the increase is not that intensive.

The OH emission from water has previously been studied by several groups using different techniques. The reported emission cross sections for OH (A $^2\Sigma^+$ – X $^2\Pi$) differ markedly. Müller et al. (1993) report a value measured at an electron energy of 100 eV that is approximately 30% lower than the value reported by Beenakker et al. (1974). However, it is necessary to note the difference between their experimental techniques. While Beenakker et al. (1974) used a cell filled with gas as a target, Müller et al. (1993) and Makarov et al. (2004) both used crossed beam set-ups to determine the cross sections. All three experiments used simple electron guns to generate electron beams. Both Beenakker et al. (1974) and Makarov et al. (2004) measure velocity dependent cross sections with a monochromator, thus requiring extrapolation from a fixed, small range in wavelength to the entire OH (A $^2\Sigma^+$ – X $^2\Pi$) band to determine absolute emission cross sections (similar to our method). According to Becker et al. (1980) the polarization of the emission is 5.2 ± 1.1 % at 11.9 eV and less than a few percent above 20 eV and Vroom & De Heer (1969) detected no polarization above 50 eV. From all studies, Shappe & Urban (2006) and Müller et al. (1993)

were the only groups that separated various vibrational bands and determined corresponding OH (A $^2\Sigma^+$ – X $^2\Pi$) cross sections.

For the reaction channel leading to OH (A $^2\Sigma^+$ – X $^2\Pi$) emission we find a threshold of 9.4 ± 0.3 eV, just above the thermochemical minimum energy (9.24 eV) and in good agreement with previous measurements (Tables 3 and 4). The emission cross section peaks around 19.5 eV, then decreases until 65 eV. Our results indicate a slight increase again with increasing impact energy, which was not expected based on the measurements by Müller et al. (1993) and Beenakker et al. (1974). We attribute this different behaviour on the heating of our set up. Khodorkovskii et al. (2009) investigated the effect of gas temperature on the electron impact induced emission cross section of OH (A $^2\Sigma^+$ – X $^2\Pi$). Their results agree with the first two papers at low nozzle temperatures (16 – 24°C), and resemble ours at nozzle temperatures of 50°C. The difference is explained by the dissociation of singlet states that are excited at these higher temperatures. Heating of our system to 60°C was necessary to avoid disruptive condensation of water vapor in the nozzle system.

### 3.3 Emission from OH+

For OH+ (A $^3\Pi$ – X $^3\Sigma^-$) we determined the cross section by integrating the surface area under the experimentally acquired spectrum (with calibrated intensity according to Sec. 2.2) between 333 and 378 nm, rather than by developing a spectral model (see Table 6 for exact integration boundaries for individual cross sections). These values bear higher uncertainties than those of OH as the experimental spectrum contains noise and possibly a small contribution from the OH (A $^2\Sigma^+$ – X $^2\Pi$). As is shown in Fig. 4, the region of the OH+ (A $^3\Pi$ – X $^3\Sigma^-$) emission also contains two excited emission features and the wavelength bands containing these features were excluded from the cross section integration. This adds to the uncertainty of our cross sections. The first feature is a strong, broad band between 342.9 – 346.5 nm which Müller et al. (1993) tentatively attributed to OH$^{2+}$. Apart from OH+ and $H_2O^+$ (next section), mass spectroscopy experiments indicate that electron impact produces a multitude of fragment ions including H+, $H_2^+$, O+, and O$^{2+}$; OH$^{2+}$ does not appear to be a major dissociation product (King & Price, 2008), suggesting that the emission is produced by one of the other neutral or ionic fragments produced by the reaction. The second feature found around 377.9 nm might be attributed to excited $O^+$ ion, but if that were the case, we would expect multiple strong emission features between 400 – 480 nm due to the $O^+$ 3p-3s transitions, which were not observed.

The energy dependent relative emission cross sections for $OH^+$ (A $^3\Pi$ – X $^3\Sigma^-$) measured at 395.5 nm is shown in the Fig. 5b. We determined a threshold around 23 ± 0.3 eV. Using an OH ionization energy of 13.017 eV (Ruscic et al. 2002) we find a thermochemical minimum appearance energy of 21.7 eV for production of excited $OH^+$ (A $^3\Pi$). We found no other reported value for this threshold, but our results are in excellent agreement with studies of the photoionization threshold of $H_2O$ leading to the production of $OH^+$ in the ground state (Ruscic et al. 2002).

After a steep increase between 23 and 50 eV, the emission cross section of $OH^+$ (A $^3\Pi$ – X $^3\Sigma^-$) reaches a plateau of and then seems to decrease slightly with increasing energy up to 100 eV. Absolute emission cross sections are given in Table 5 and add up to a total of $6.4 \times 10^{-20}$ cm$^2$, which is a factor of 25 less than the emission from neutral OH (A $^2\Sigma^+$ – X $^2\Pi$). Our results are significantly lower than those reported by Mueller et al. (1993) whose separate emission band cross sections add up to a total emission cross section of $29.8 \times 10^{-20}$ cm$^2$.

*3.4 Emission from $H_2O^+$*

In the spectral region between 430 and 750 nm more than 400 sub bands of $H_2O^+$ (A $^2A_1$ – X $^2B_1$) transition can be found (Kuchenev & Smirnov 1996). In Figure 7, a detail of part of this spectral region (between 430 and 560 nm) is shown. The sub-bands were identified according to Kuchenev & Smirnov (1996). To determine emission cross sections a spectral resolution much higher than possible with our spectrometer is necessary as the bands are spread over a large spectral range and overlap each other.

We measured the energy dependence of three relatively bright $H_2O^+$ features that were reasonably separated from other spectral features with peaks at 496.3 nm, 503 nm, and 519.75 nm (Fig. 8). These features are all part of the $H_2O^+$ (A $^2A_1$ – X $^2B_1$) band, corresponding to transitions (0,12,0) – (0,0,0) - p branch; (0,13,0) – (0,1,0) - p and r branches; and (0,11,0) – (0,0,0) - p branch respectively.

In Figure 8 the relative cross section curves for these three transitions are shown. The shapes of the curves are similar to each other and also to the curve reported by Müller et al. (1993) for the transition (0,16,0) – (0,2,0). We find a threshold value of 15.8 ± 0.3 eV, in agreement with the thermochemical minimum of 14.62 eV and with the measured threshold value of 15.0 ± 0.5 reported by Müller et al. (1993).

In Table 6 the resulting cross sections at 50 eV incident electron energy are shown including the integration boundaries used for their determination. The values are in reasonable agreement with Kuchenev & Smirnov (1996) who show values approximately in the range of $1 \times 10^{-21}$ cm$^2$ and $4 \times 10^{-20}$ cm$^2$ for individual bands at 50 eV electron impact energy. However, we note that due to the insufficient optical resolution of our experiment the values contain signal from several transitions. For comparison, Müller et al. (1993) only give a lower limit of the cross section value for the spectral region 350 – 500 nm at 100 eV which is $>1.25 \times 10^{-18}$ cm$^2$.

## 4. EXAMPLES OF DIAGNOSTIC APPLICATIONS IN COMETS

Dissociative electron impact excitation provides a remote diagnostic of both the neutral gas and the electrons interacting with it (Section 1), and water vapor is ubiquitous in our solar system. Through spectral analysis and modelling that implements dissociative cross-sections as a function of electron energy the dominant emission source can be identified, and if spectral signal is high enough to capture a wide range of emission features with varying threshold energies it may even be possible to remotely determine plasma properties. However, to fully employ its diagnostic qualities, we need to distinguish between different mechanisms that excite fragment species (Feldman 2004). In comet atmospheres, those are mainly resonant fluorescent excitation (Swings 1941; Schleicher & A'Hearn 1988) and emissive photodissociation ('prompt' emission; Bertaux 1986; Budzien & Feldman 1991; A'Hearn et al. 2015; La Forgia et al. 2017*)*. In this section, we will briefly highlight how the emission from electron impact on water vapor in Near-UV optical wavelengths differs from the emission from these two other processes. We will only compare the spectral signatures; it is of note that in comet atmospheres the morphology of emission due to direct excitation mechanisms (i.e. emissive photodissociation and dissociative electron impact emission) maps the distribution of parent species (H$_2$O here) and can be found closer to the nucleus, whereas fluorescent emission maps the fragment species resulting in a flatter, more extended distribution (cf. Bertaux 1986; Combi et al. 2004; Bodewits et al. 2016). We also note that electron impact UV emission has been evaluated in several protoplanetary disks (France et al. 2011); improved cross sections for further molecular species may allow remote characterizations of plasma environments that are beyond the reach of current spacecraft capabilities.

First, we compare fluorescent excitation and dissociative electron impact excitation of the Hydrogen Balmer series. While H$_\alpha$ has been observed in some comets (e.g. Combi et al. 1999; Shih et al. 1985; Cochran & Cochran 2002), the observations are challenging owing the low surface brightness of fluorescence emission from the million-km-sized Hydrogen envelope of comets, and because it can be easily overwhelmed by geocoronal emission. Based on the results of the *Rosetta* mission (Bodewits et al. 2016), we expect electron impact emission to dominate in the inner coma (~100 km) of comets with low production levels (~10$^{27}$ molecules s$^{-1}$).

We estimated excitation rates of atomic hydrogen (cf. Feldman et al. 2004) exposed to Sun light using oscillator strengths from Wiese & Fuhr (2009). For the solar irradiance of the Lyman lines we used high-resolution solar spectrum acquired by the Extreme Ultraviolet Monitor on board the *Maven* spacecraft (Eparvier et al. 2015) during the solar maximum in 2014 and we assumed a heliocentric distance of 1 au and a heliocentric velocity of 0 km/s. To determine the line strength of the Balmer series, one needs to consider the cascade population of states below those initially populated. However, because the excitation of states with n > 4 is very small compared to n = 3, 4 (factor 20 smaller), and because we only excite states with angular momentum l = 1 we can ignore cascade population here. This implies that for the fluorescence emission, the line strength is given by the excitation rate weighed by the transitions branching ratios, which are 0.88 for the Lyman series and 0.12 for the Balmer series (Omidvar 1980). The resulting relative line strengths are given in Table 7. The ratio between the H$_\alpha$ and H$_\beta$ is similar for electron impact (above energies of 40 eV) and resonant fluorescence, but electron impact results in more higher-order (n > 4) Balmer lines. In the case of electron impact, these ratios do vary with energy (Fig. 3b); they increase steeply at energies exceeding the onset of the channels and reach a maximum at electron impact energies of approximately 40 eV, after which they decrease slightly again. This implies that the ratio between the $H_\alpha$, $H_\beta$, and $H_\gamma$ can be used as a remote diagnostic of the electron temperature.

Second, we consider the spectrum of OH (A $^2\Sigma^+$ – X $^2\Pi$). As is shown in Fig. 9, the three different excitation processes lead to distinctly different emission spectra between 260 – 335 nm. Emissive photodissociation by Ly-$\alpha$ emission is known to produce OH (A $^2\Sigma^+$) with relative populations in the $v$ = 0, 1, and 2 of 1:0.3:≤0.01 (Carrington et al. 1964; Harich et al. 2000; A'Hearn et al. 2015). It also leads to the population of high rotational states, which typical quantum numbers 17 ≤ $N(v=0)$ ≤ 22 and 12 ≤ $N(v=1)$ ≤ 17 for the two vibrational levels. The population of vibrational levels by fluorescent excitation will depend heavily on the heliocentric velocity of a comet through the Swings effect, but generally leads to the population of much lower rotational states ($N$ ≤ 5; Schleicher & A'Hearn 1988; La Forgia et al. 2017). Finally, dissociative electron impact leads to a relatively large population of the $v$ = 2 level (1:0.3:0.2) for $v$ = 0, 1, and 2, respectively, and to a superposition of two clearly non-thermal distributions of the rotational states (Fig. 6). The resulting OH (A $^2\Sigma^+$ – X $^2\Pi$) spectra produced by fluorescence excitation, emissive photodissociation, and electron impact dissociation excitation are markedly different (Fig. 9). In the fluorescence spectrum, the emission peaks between 305 – 310 nm (Schleicher et al. 1988). Emissive photodissociation results in a spectrum with much stronger emission between 280 and 290 nm, additional emission at 307 nm, and two broad maxima in the region between 305 and 310 nm and 312-318 nm (La Forgia et al. 2017). Electron impact dissociation results in a more continuous spectrum with maxima at 305, 309, 312, and 315 nm.

Third, the emission features discussed here all correspond to different reaction channels, each with its own threshold (Tables 3 and 4). Whereas in astrophysical environments electrons will not be mono-energetic as in our set up (cf. Broiles et al. 2016), the onset and disappearance of emission features is tied to the temperature of the incidence electrons. For example, the OH (A $^2\Sigma^+$ – X $^2\Pi$) band occurs around 9.2 eV, or about 100,000 K; $H\alpha$ appears at 17.7 eV, or 200,000 K. Optical and Near-UV emission features of electron impact dissociation can thus be used as a remote plasma diagnostic.

Fourth, the electron impact induced spectra can be used as templates to identify emission lines in astrophysical spectra. For example, while analysing our results, we compared the $H_2O^+$ ($A^2A_1 - X^2B_1$) spectra from our experiments and those by Kuchenev & Smirnov (1996) with observations of comet ion tails (Wyckoff et al. 1999; Kawakita & Watanabe 2002) and found that many unidentified lines could be attributed to transitions from the higher vibrational levels of the $H_2O^+$ ($A\ ^2A_1 - X\ ^2B_1$) band, levels which were not considered in fluorescence models.

To facilitate the application of our results in astrophysical models, we provide polynomial fits to the data in Appendix A. These polynomials are purely mathematical fits with no theoretical basis. The determined cross section curves (both absolutely calibrated and relative ones) were smoothed and fitted by polynomial functions of the 9th order valid in the range from the threshold of the process up to 100 eV. In specific cases (H$\beta$ and H$\gamma$) it was necessary to divide the range into two parts and fit them separately since using only one function for the whole range would lead to discrepancies between fit and real values around the second threshold.

## 5. CONCLUSIONS

Electron impact reactions can be used to remotely detect and characterize volatiles in the solar system. To discover and interpret this emission, a thorough understanding of dissociative electron impact reactions is required. Water is ubiquitous in the solar system and one of the main constituents of the tenuous atmospheres of small bodies such as comets. We have experimentally studied the dissociative excitation of water molecules using a crossed-beam set up equipped with multiple electron sources and spectrometers. This allowed us to characterize the emission cross sections of excited OH, atomic hydrogen, and the ions $H_2O^+$ and $OH^+$, as well as the impact energies above which those excited products were formed. The first emission features to appear are the OH ($A\ ^2\Sigma^+ - X\ ^2\Pi$) bands between 260 – 335 nm, with a threshold of 9.4 ± 0.3 eV. The $OH^+$ ($A\ ^3\Pi - X\ ^3\Sigma^-$) and $H_2O^+$ ($A\ ^2A_1 - X\ ^2B_1$) features occur next at incident energies above 13 ± 0.3 and 15.8 ± 0.3 eV, respectively. Finally, above 17.7 ± 0.3 eV, the Hydrogen Balmer series was detected. These values are in good agreement with thermochemical estimates and previous experiments.

We have investigated the brightest features, the OH (A $^2\Sigma^+$ – X $^2\Pi$) and the Hydrogen Balmer series in the most detail, comparing our results to previously published studies. Theoretical simulations of the OH (A $^2\Sigma^+$ – X $^2\Pi$) bands were used to determine emission cross sections of individual vibrational transitions which are not distinguishable in experimental spectrum. To aid the diagnostic application of our experimental results, we produced empirical fits to the electron impact dependent emission cross sections.

Following the findings of the *Rosetta* mission to comet 67P/Churyumov-Gerasimenko that electron impact reactions might rival photolysis and fluorescence in certain circumstances, we explored the diagnostic application of electron impact excitation in cometary atmospheres. For hydroxyl, electron impact dissociation leads to a ro-vibrational excitation that is distinctly different than that produced by resonant photo-fluorescence and emissive photodissociation. This phenomenon can be used for remote determination of the primary process leading to excitation. While our experiment cannot be used to reconstruct the detailed distribution over the quantum and angular momentum states of the excited Hydrogen, the relative intensities of the Hydrogen Balmer series are very different when produced by electron impact reactions with $H_2O$ rather than through the photo-fluorescence of atomic hydrogen. In short, given the proper signal to noise, spectral observations of interactions between neutral molecules and energetic plasmas, like those within cometary comae, can isolate portions of the tenuous gas dominated by electron impact or photofluoresence based on the relative intensities of the Balmer series.

While water is the main volatile in comets, their comae contain many other molecules such as $CO_2$, CO, HCN, NH, and $O_2$, species that are also present in the atmospheres of other small bodies in the solar system. We hope to explore the electron induced dissociative excitation of these molecules and their fragments and ions in future studies.

Acknowledgements: We thank Dr. D. Schleicher (Lowell Observatory) for sharing the results of his OH fluorescence calculations, and Dr. H. Kawakita for discussing the line identification of $H_2O^+$ in comet tails.  Support for Program number HST-GO-15625.001 (DB, JN) was provided by NASA through a grant from the Space Telescope Science Institute, which is operated by the Association of Universities for Research in Astronomy, Incorporated, under NASA contract NAS5-26555. This project has received funding from the European Union's Horizon 2020 research and innovation programme under grant agreement No 692335. This work was supported by the Slovak Research and Development Agency APVV-15-0580 and the Slovak Grant Agency for Science (contract no. VEGA 1/0733/17).

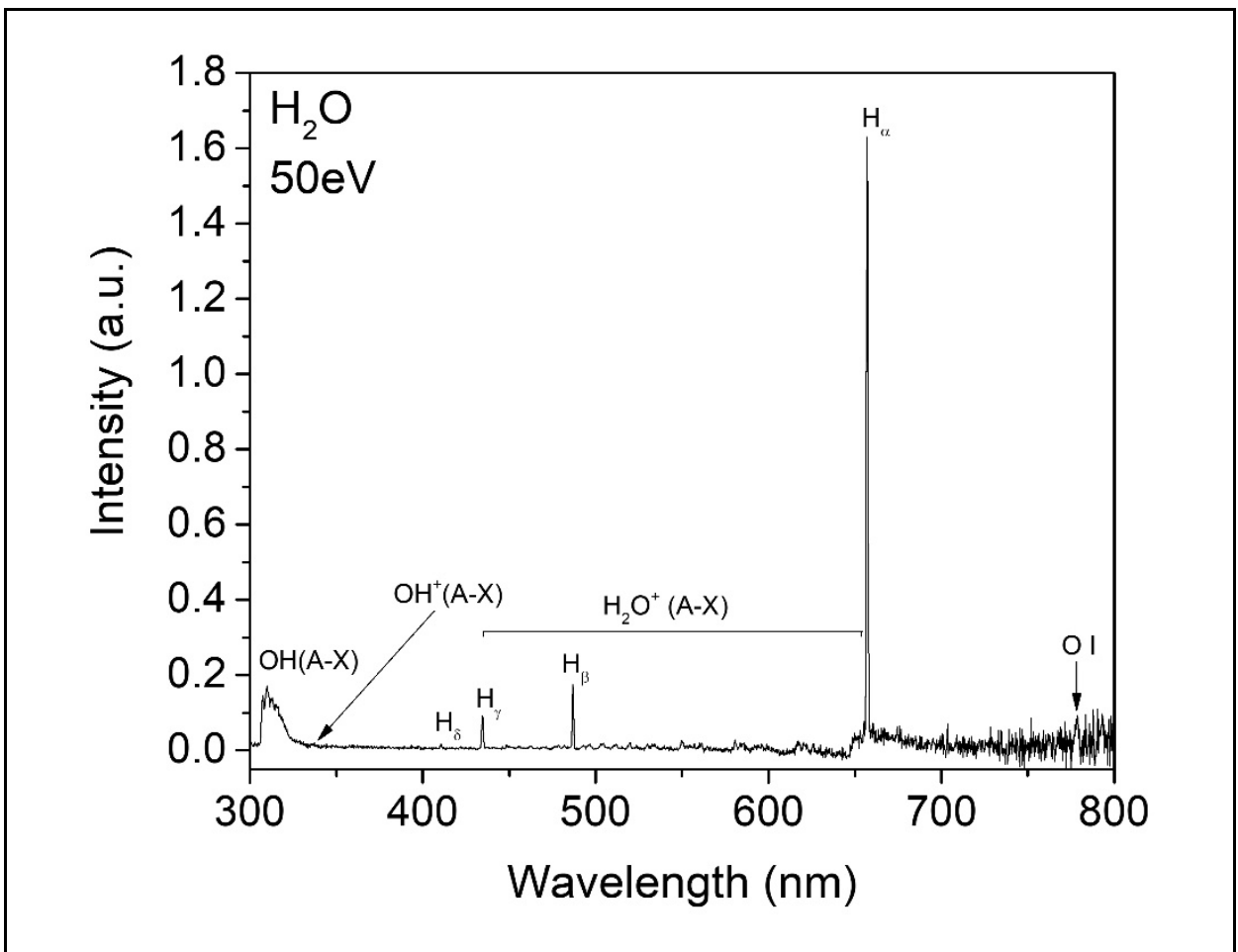

**Fig. 1** - Overview of the emission spectrum and the main features induced by electron impact on water vapor at an incident energy of 50 eV. Corrected for instrument sensitivity.

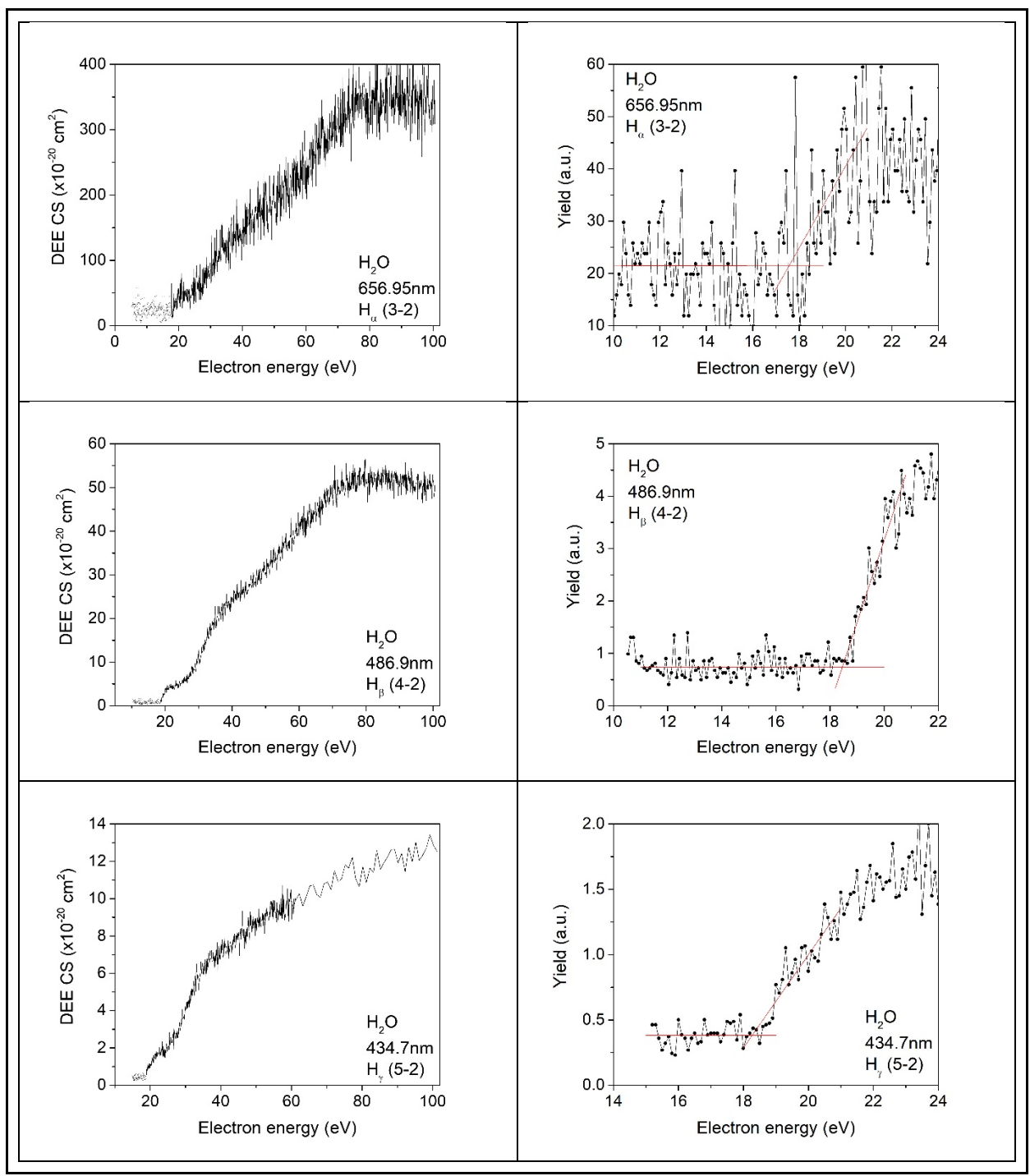

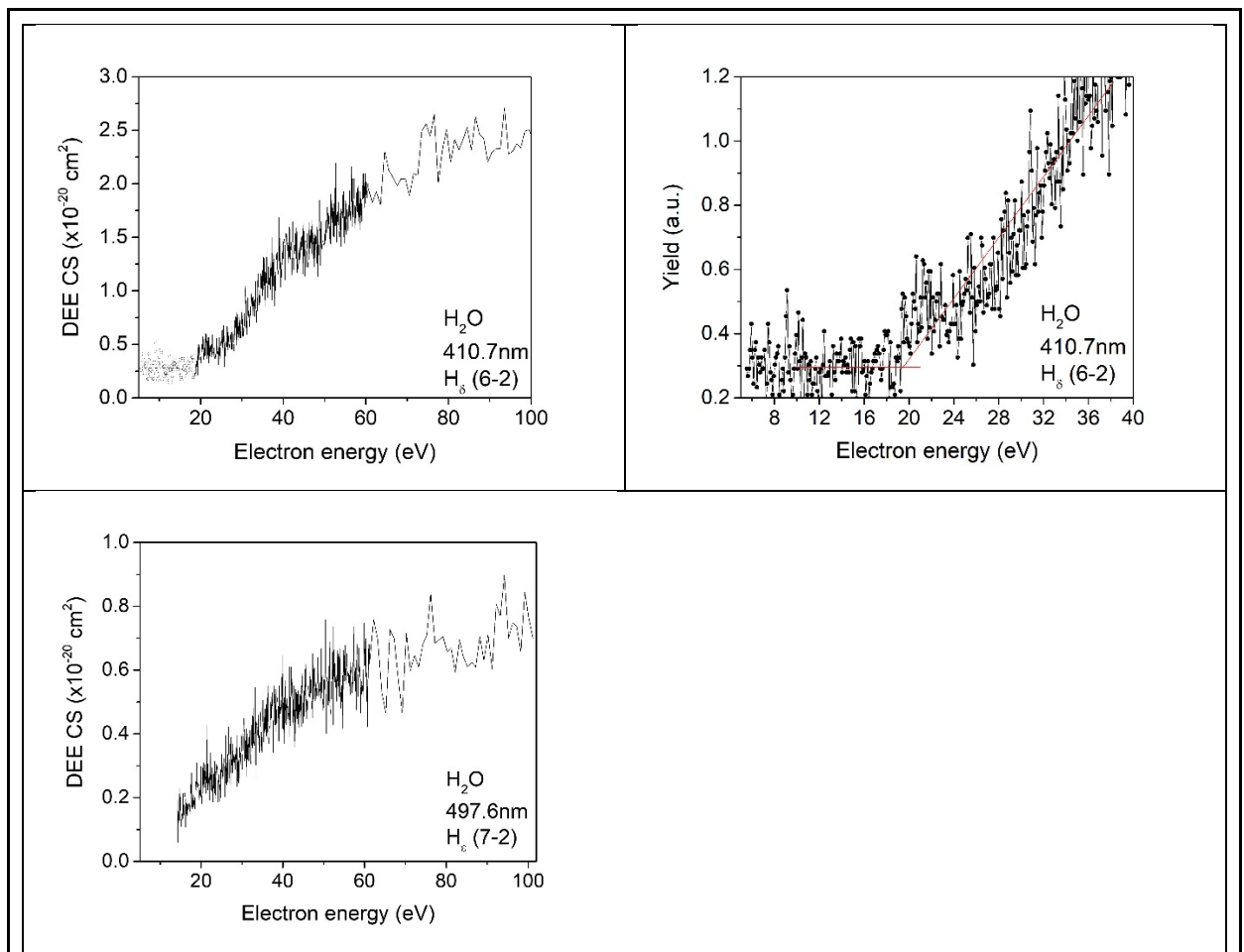

**Fig. 2** – Electron energy dependence of emission cross sections of the Balmer series ($H_\alpha$, $H_\beta$, $H_\gamma$, $H_\delta$, and $H_\varepsilon$ (left column) and corresponding fits of the thresholds (right column). The signal-to-noise ratio of the $H_\varepsilon$ emission line prevents fitting the thresholds.

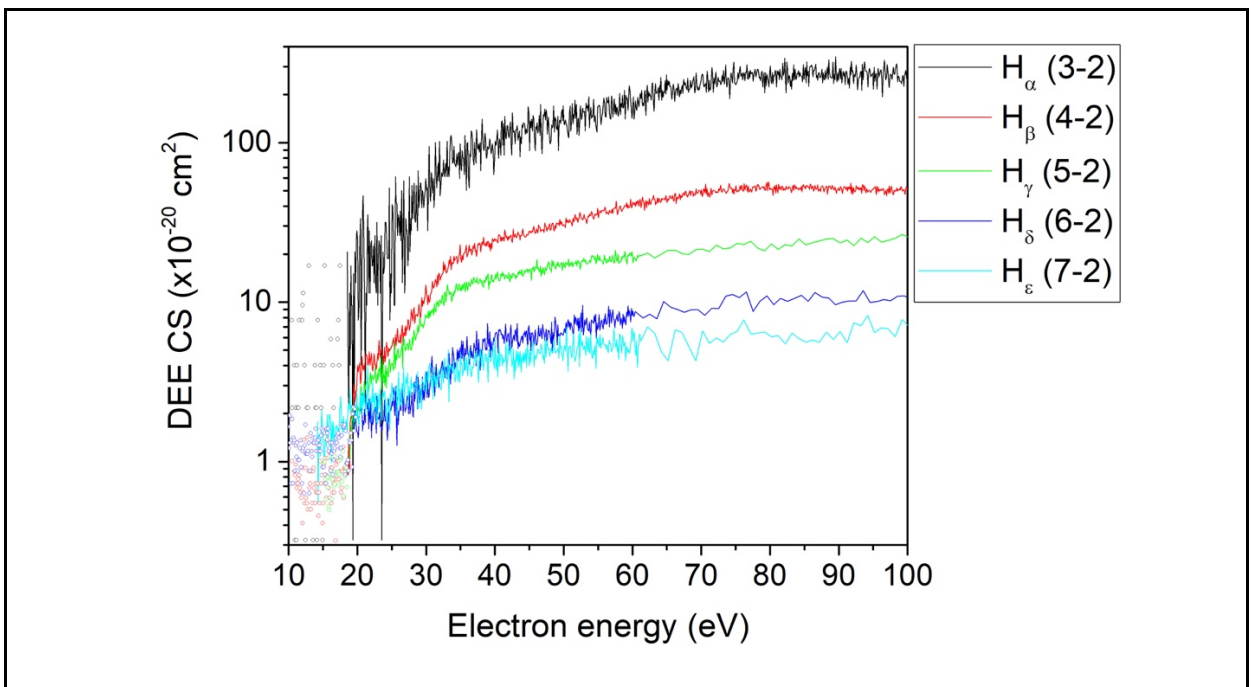

**Fig. 3a** – Comparison of emission cross sections of the Hydrogen Balmer series corrected for the loss of emission owing to the limited field of view.

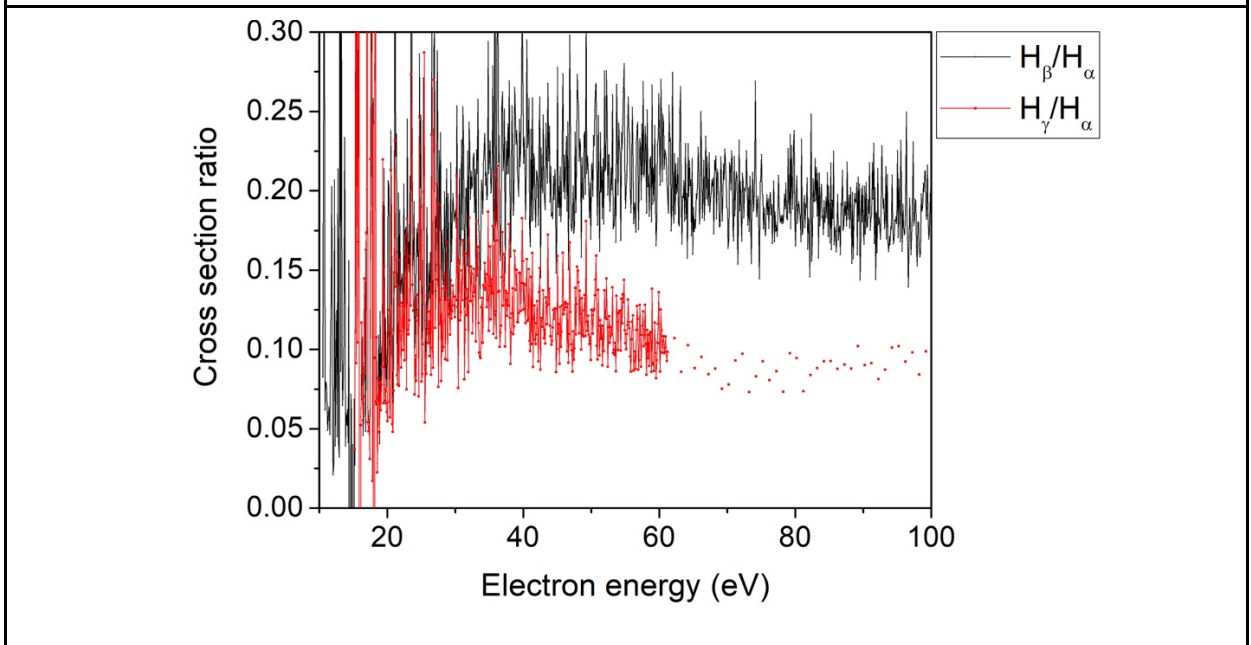

**Fig. 3b** – Line ratios of the Hydrogen Balmer series vs electron impact energy. Hβ/Hα – black line; Hγ/Hα- red line.

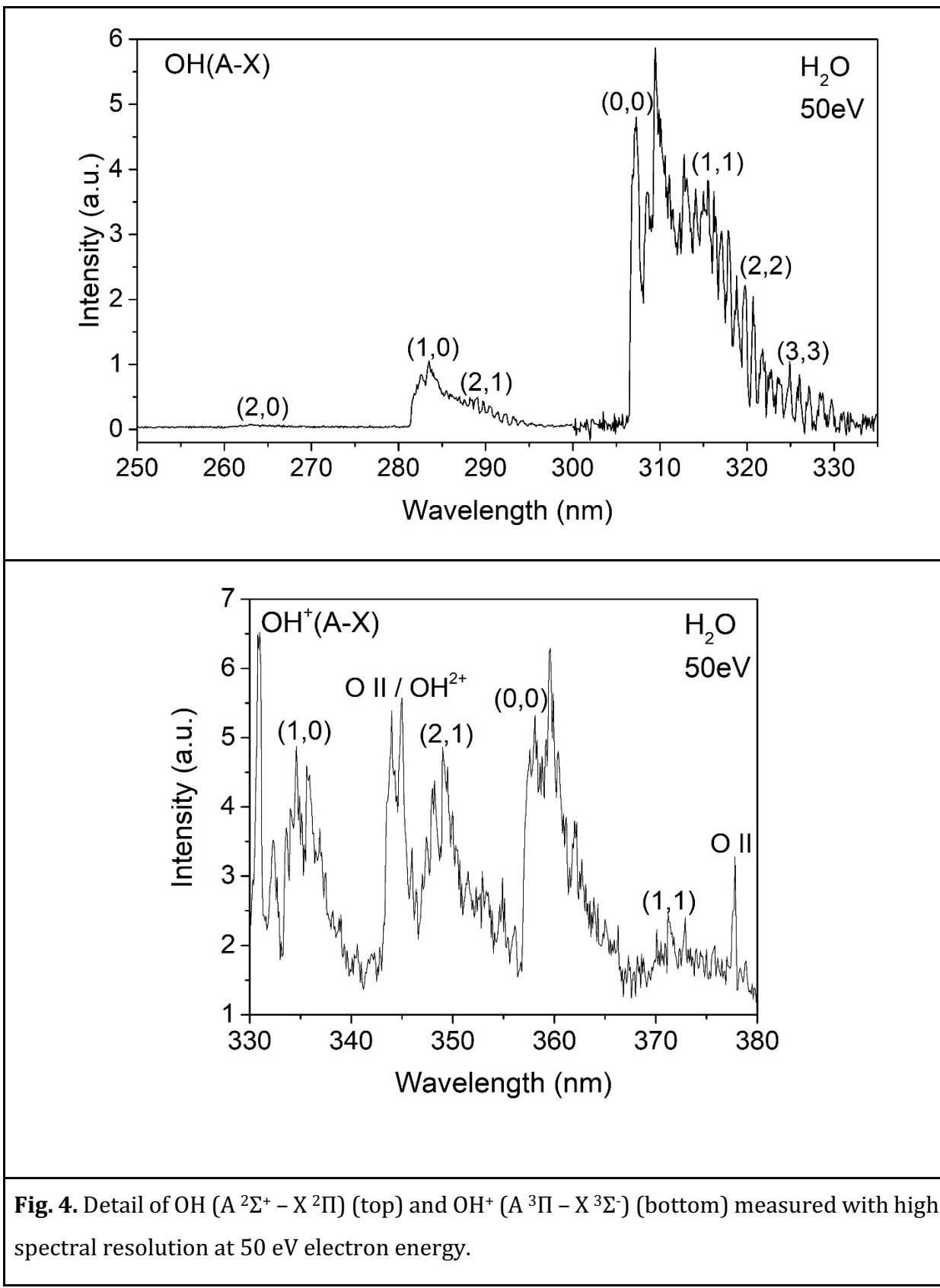

**Fig. 4.** Detail of OH (A $^2\Sigma^+$ – X $^2\Pi$) (top) and OH⁺ (A $^3\Pi$ – X $^3\Sigma^-$) (bottom) measured with high spectral resolution at 50 eV electron energy.

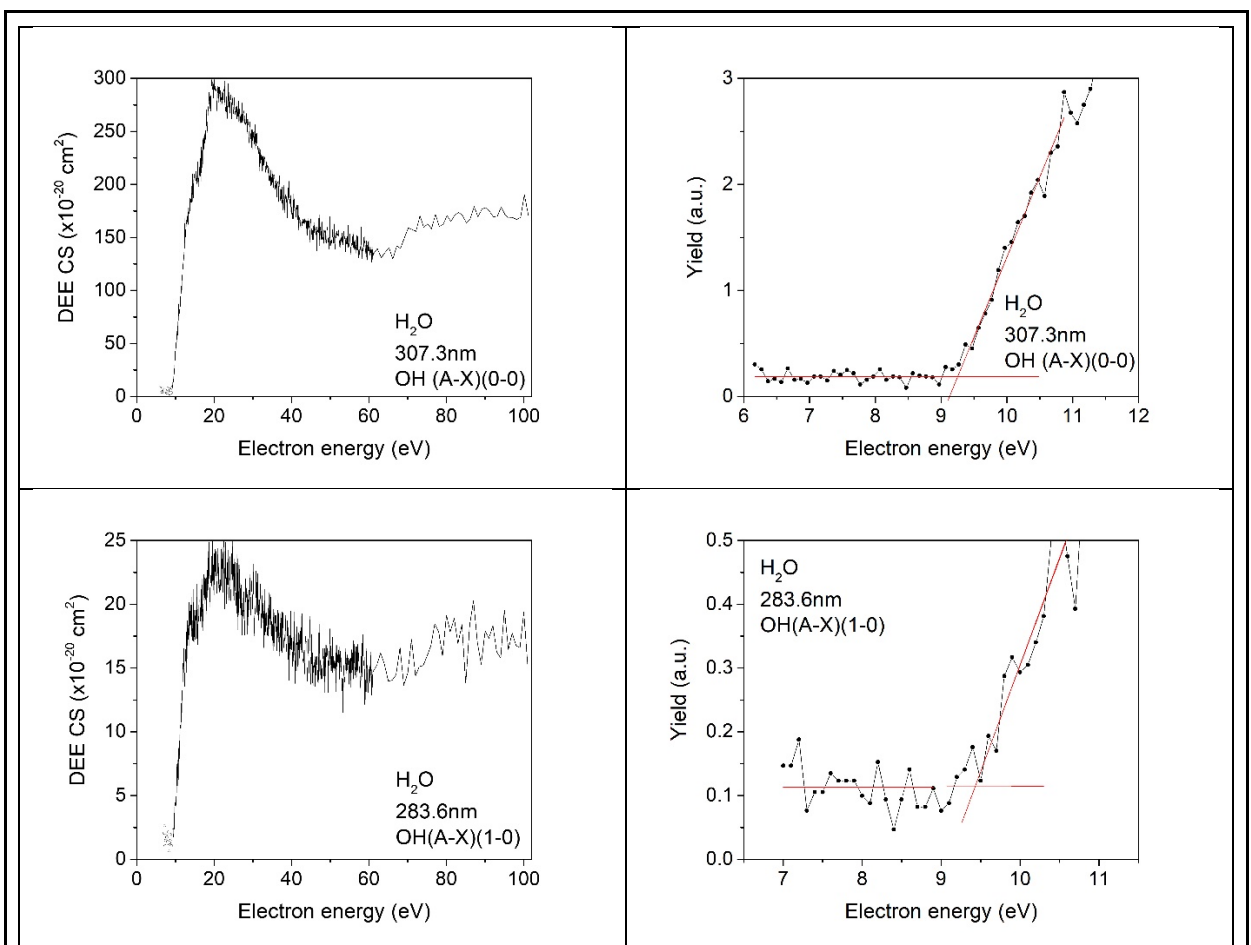

**Fig. 5a** – Electron energy dependence of emission cross sections of OH A $^2\Sigma^+$ – X $^2\Pi$ (0-0) around 307.3 nm (top left); OH A $^2\Sigma^+$ – X $^2\Pi$ (1-0) around 283.6 nm (bottom left). Figures in the right column show a close up of the cross sections around their threshold energies. Total emission cross sections are given in Table 5.

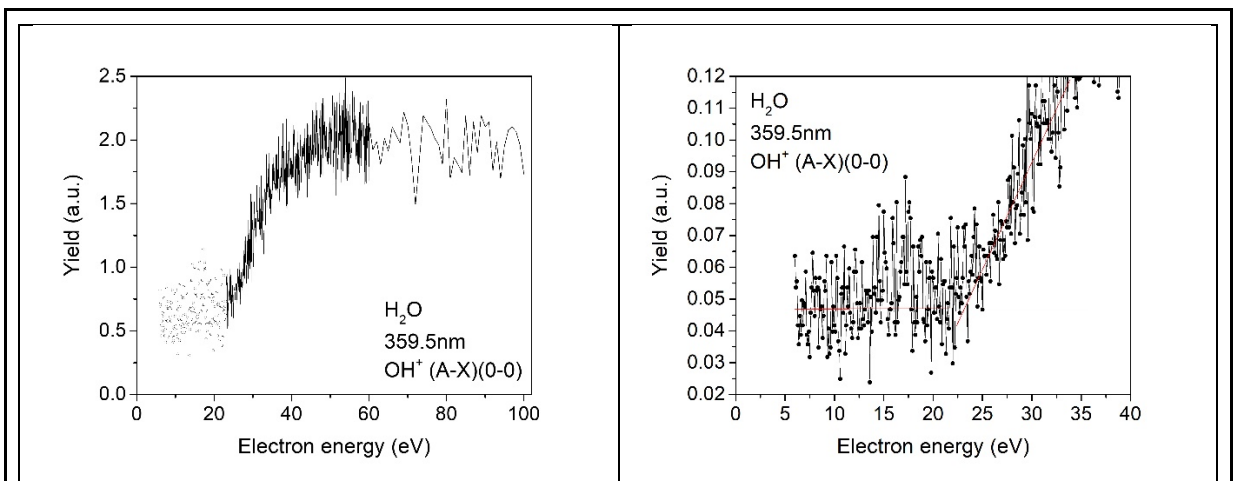

**Fig. 5b** – Electron energy dependence of relative emission cross section of $OH^+$ $A\ ^3\Pi - X\ ^3\Sigma^-$ (0-0) band around 359.6 nm. Figure in the right column shows a close up of the cross section around its threshold energy. Note that the curve is valid for part of the band only at the specific wavelength listed. Total emission cross sections are given in Table 6.

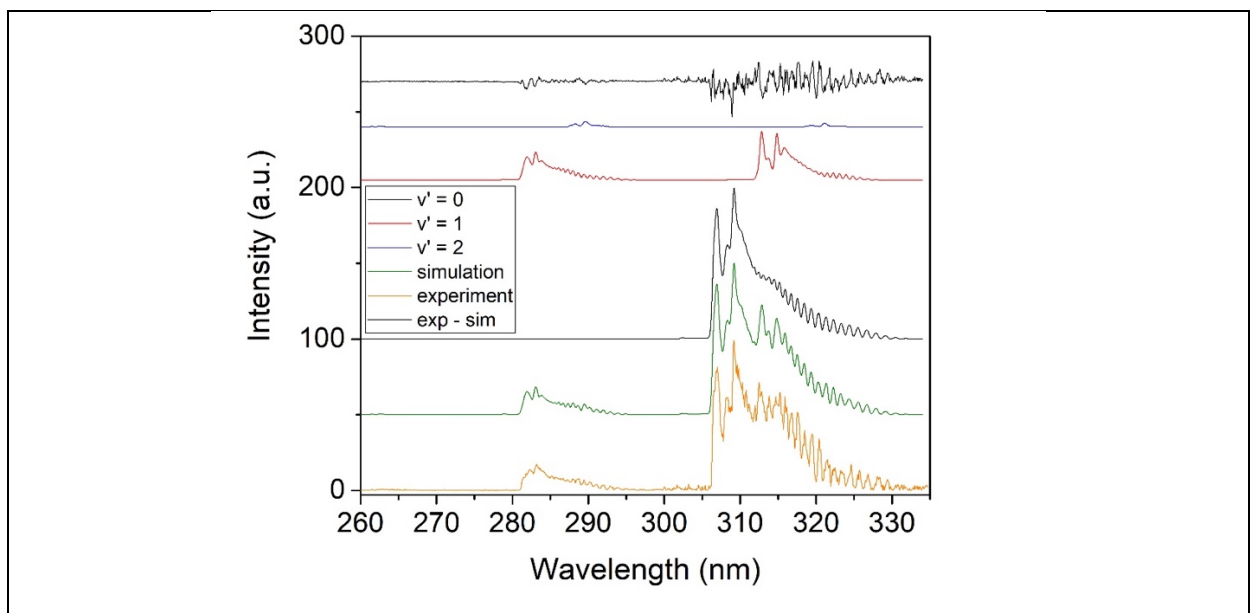

**Fig. 6a** - Comparison of the synthetic OH (A $^2\Sigma^+$ – X $^2\Pi$) spectrum (green) with the experimental spectrum (orange). Individual components of the synthetic spectrum for $v'$ = 0 (top black), $v'$ = 1 (red) and $v'$ = 2 (blue), as well as the difference between the synthetic and experimental spectrum (center; black) are also shown.

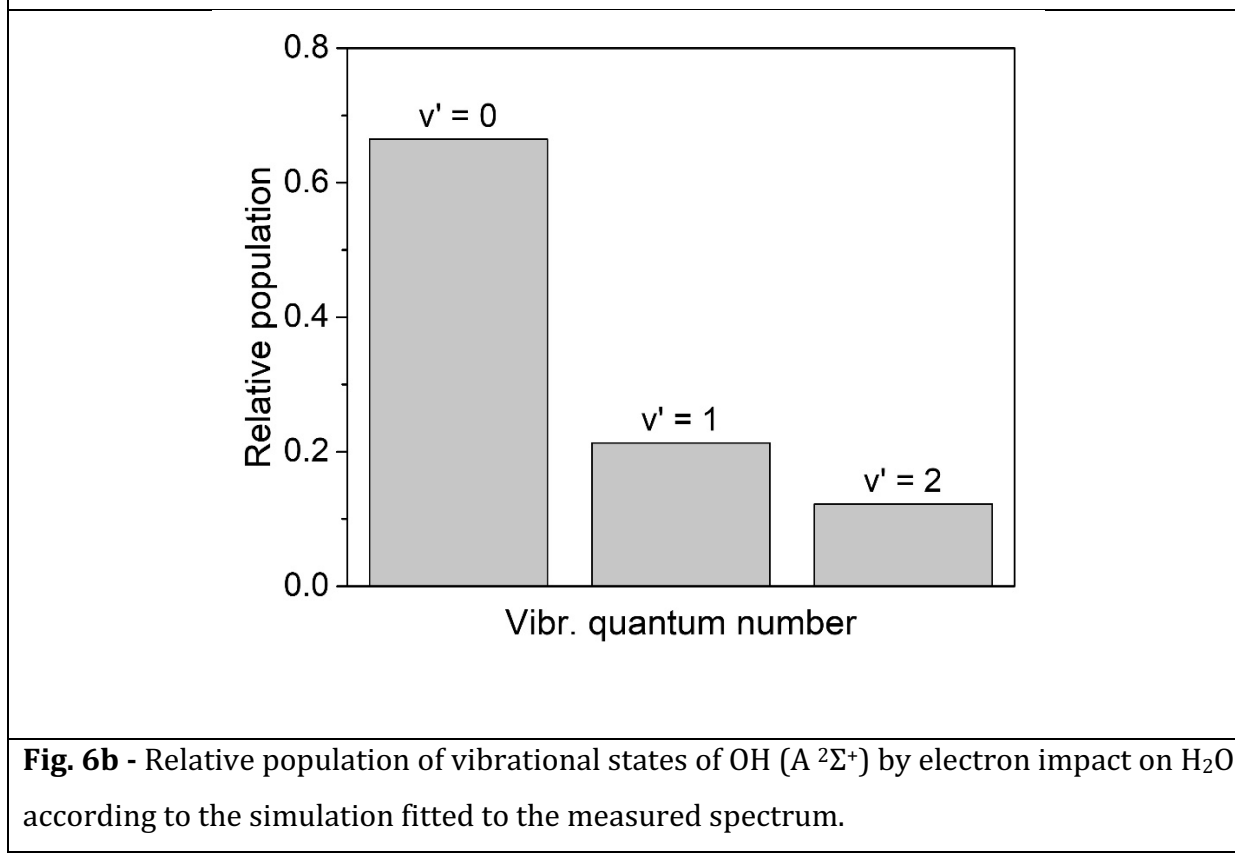

**Fig. 6b** - Relative population of vibrational states of OH (A $^2\Sigma^+$) by electron impact on $H_2O$ according to the simulation fitted to the measured spectrum.

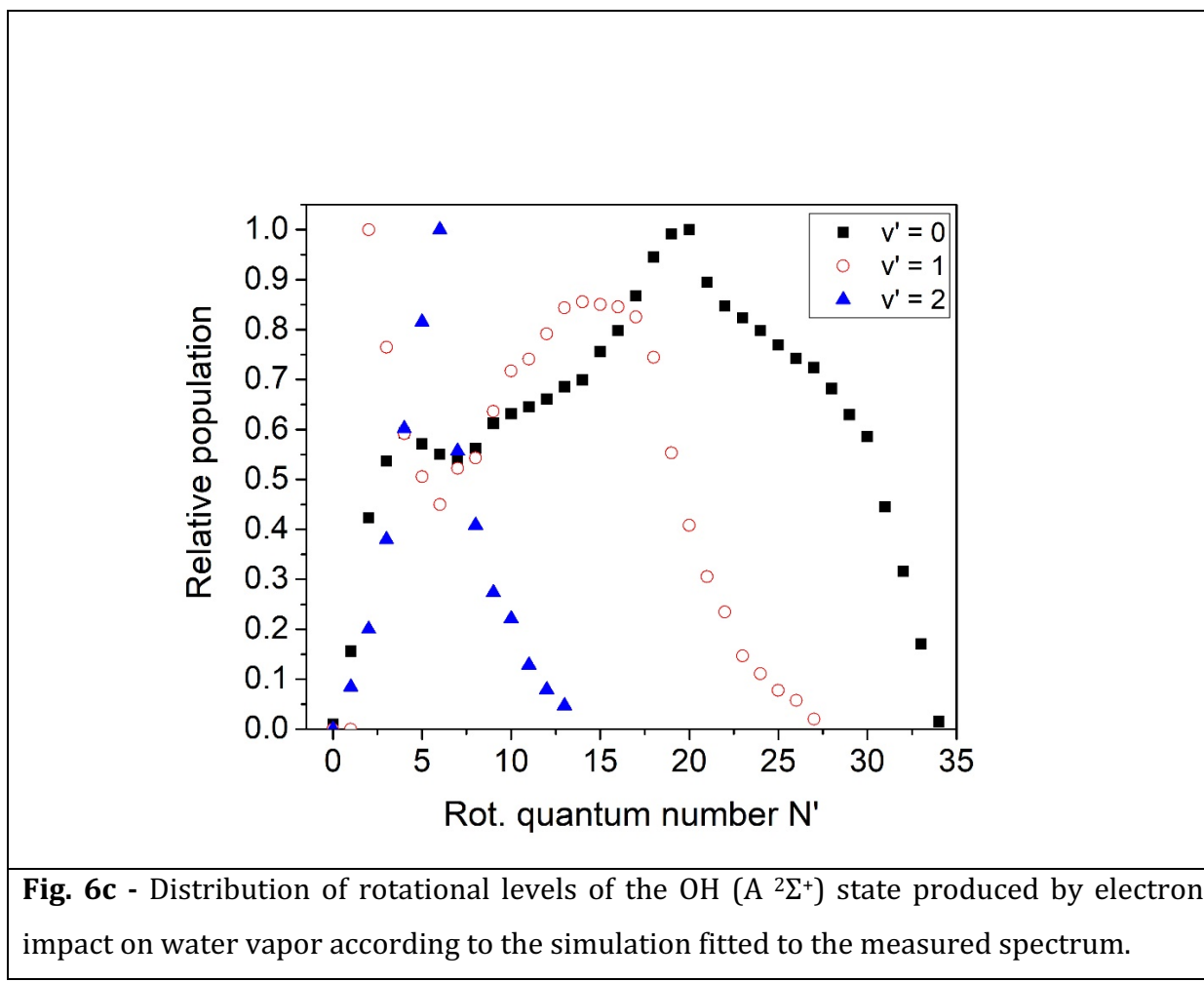

**Fig. 6c -** Distribution of rotational levels of the OH (A $^2\Sigma^+$) state produced by electron impact on water vapor according to the simulation fitted to the measured spectrum.

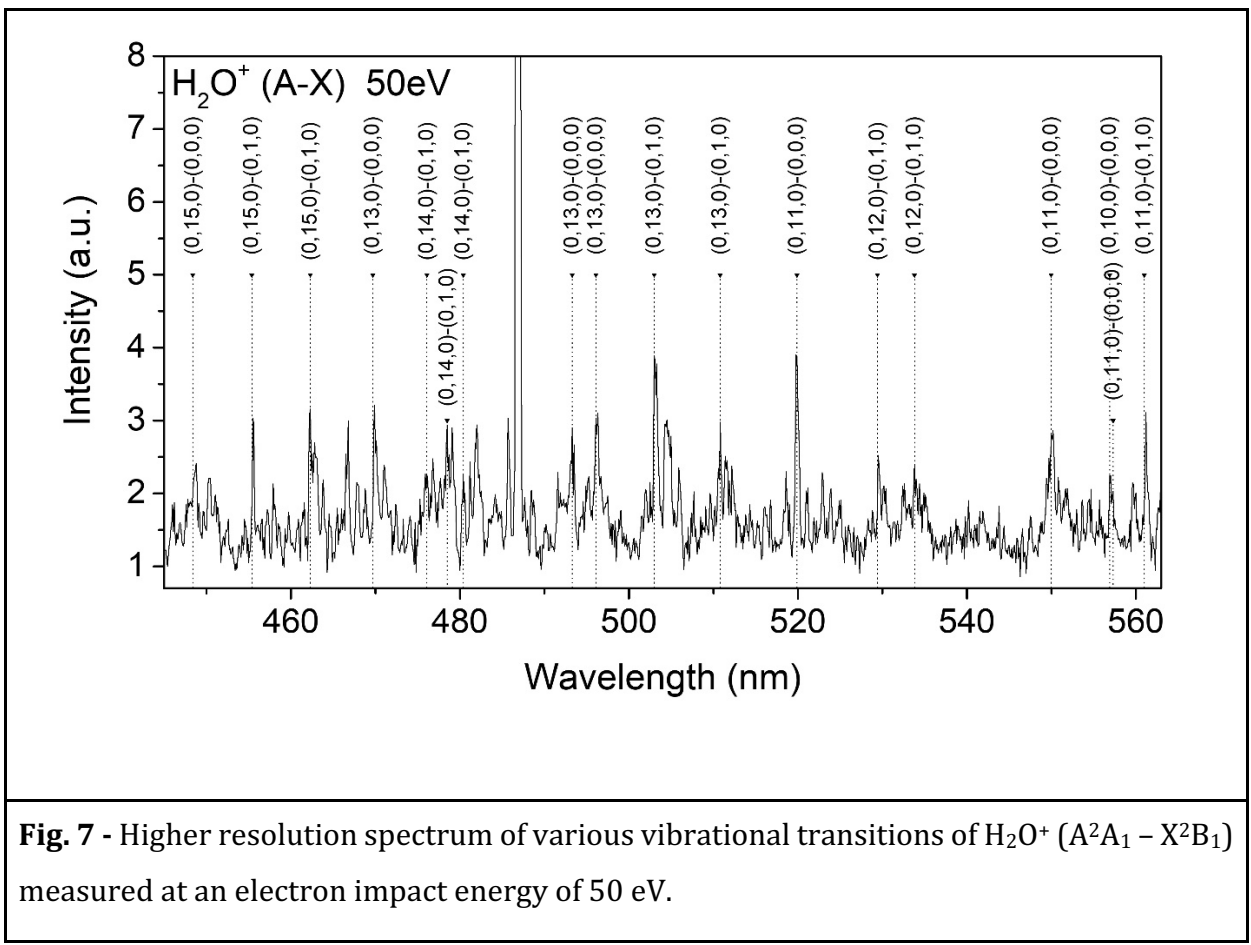

**Fig. 7 -** Higher resolution spectrum of various vibrational transitions of $H_2O^+$ ($A^2A_1 - X^2B_1$) measured at an electron impact energy of 50 eV.

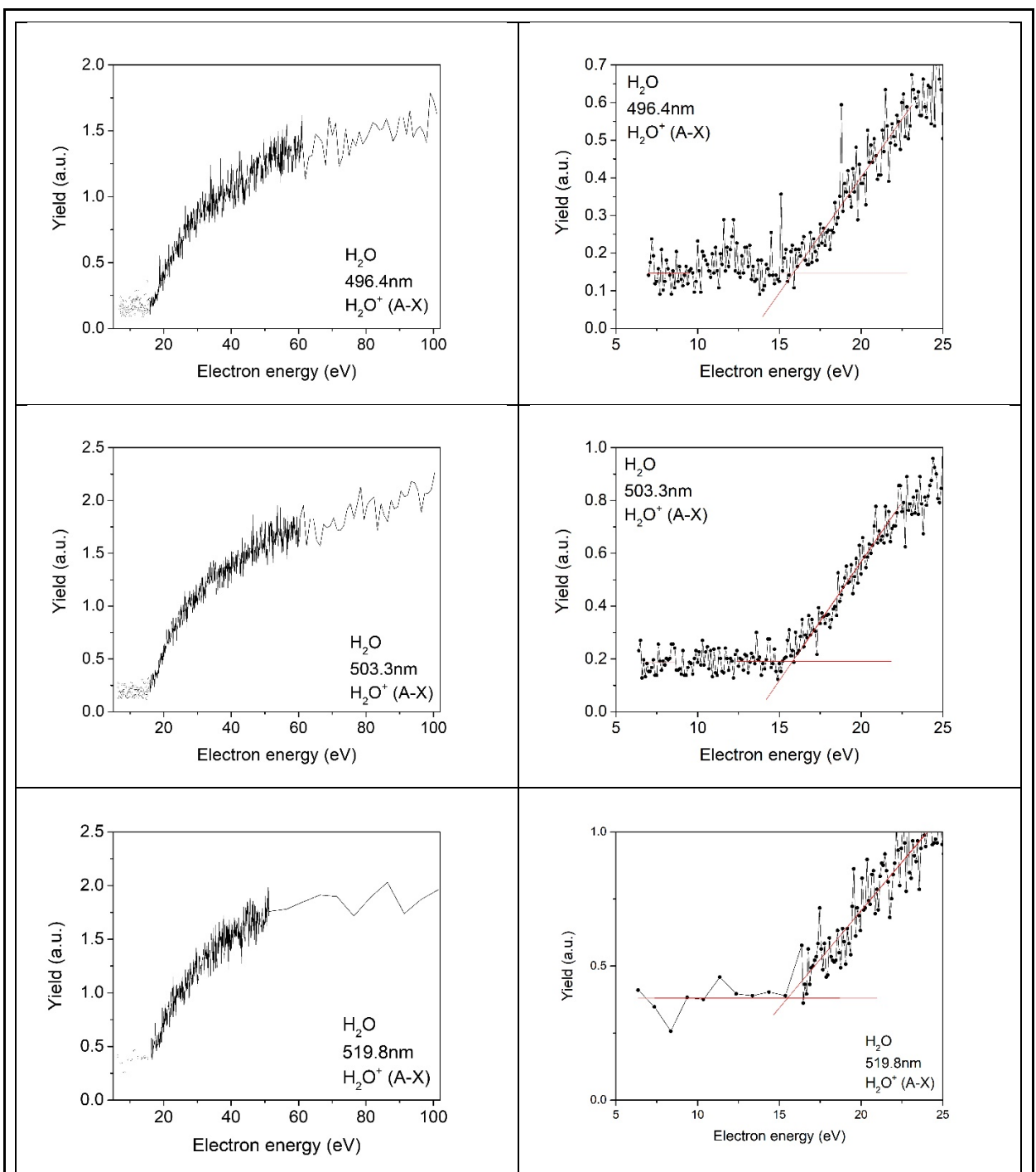

**Fig. 8** – Electron impact energy dependence of emission cross sections of different emission features that are part of the ($H_2O^+$ $A^2A_1$ – $X^2B_1$) band, measured at 496.4 nm (top left), 503.3 nm (top right), and 519.8 nm (bottom). Threshold fits are shown in the right column.

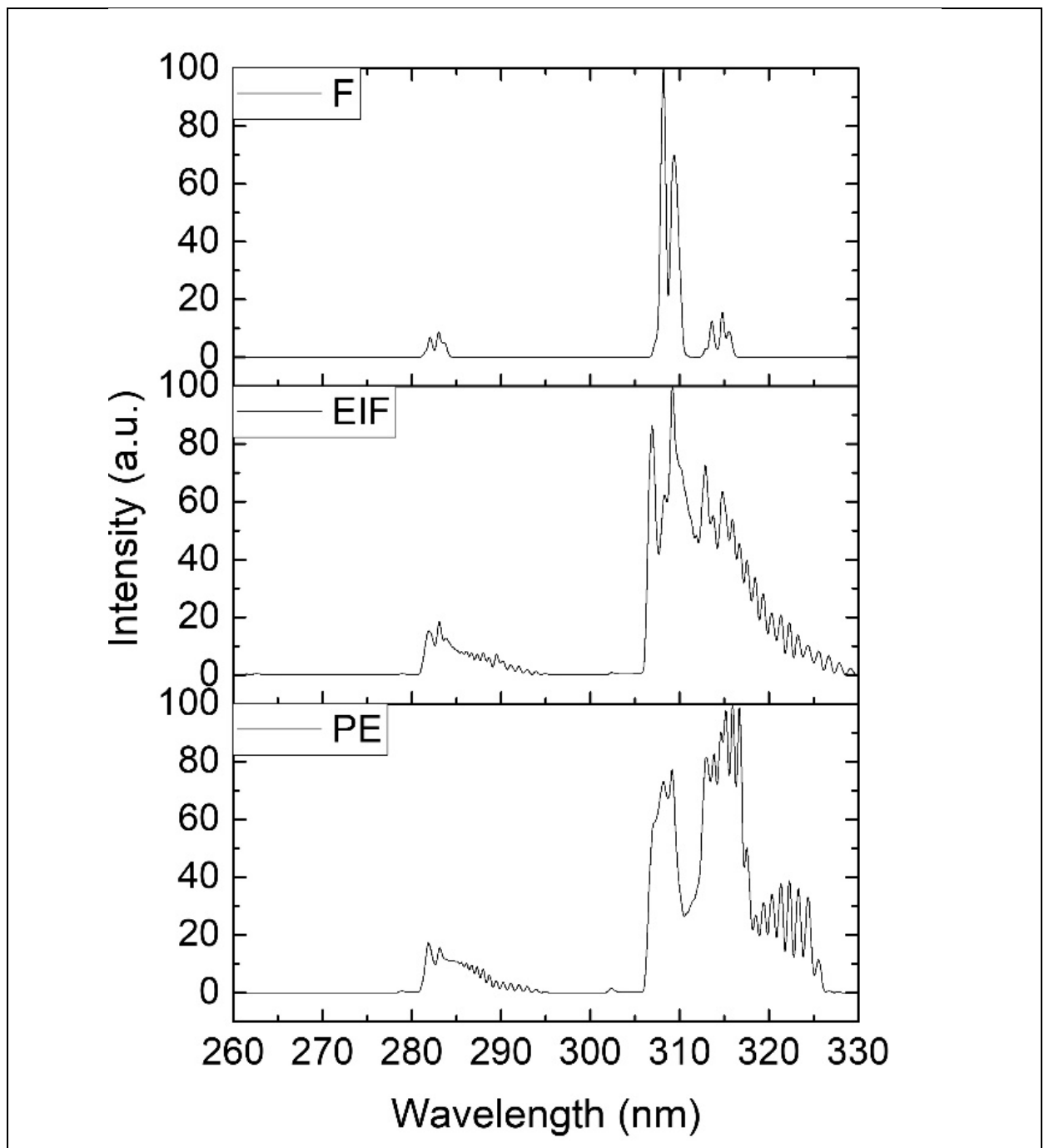

**Fig. 9** – Comparison of simulated OH ($^2\Sigma^+$ – X $^2\Pi$) spectra caused by different processes. Top: fluorescence excitation of OH at a heliocentric velocity of 2.13 km s$^{-1}$ (based on a level distribution from D. Schleicher, *priv. comm.*). Center: electron impact of H$_2$O at an incident

energy of 50 eV (this paper). Bottom: emissive photodissociation of $H_2O$ (La Forgia et al. 2017). The spectral resolution for these simulations is 0.5 nm.

**Table 1** – Lifetimes of the excited states (Wiese & Fuhr, 2009), the distance a hydrogen atom travels in that time assuming an energy of 0.26 eV (7 km/s; Makarov et al. 2004), and the fraction of the population that decays within the field of view for each emission line in the Balmer series.

| Transition | Name | Lifetime (ns) | Distance (mm) | Decayed (%) |
|---|---|---|---|---|
| 3 – 2 | $H_\alpha$ | 22.73 | 0.16 | 100 |
| 4 – 2 | $H_\beta$ | 119.1 | 0.83 | 83 |
| 5 – 2 | $H_\gamma$ | 395.3 | 2.77 | 42 |
| 6 – 2 | $H_\delta$ | 1028 | 7.19 | 19 |
| 7 – 2 | $H_\varepsilon$ | 2273 | 15.91 | 9 |

**Table 2** – Emission cross sections $\sigma$ (in units of $10^{-20}$ cm$^2$) for hydrogen Balmer series at electron impact energies of 50 and 100 eV for our measurements and previously reported values.

| $\sigma$ | $H_\alpha$ | $H_\alpha$ | $H_\beta$ | $H_\beta$ | $H_\gamma$ | $H_\gamma$ | $H_\delta$ | $H_\delta$ | $H_\varepsilon$ | $H_\varepsilon$ |
|---|---|---|---|---|---|---|---|---|---|---|
| **Energy (eV):** | 50 | 100 | 50 | 100 | 50 | 100 | 50 | 100 | 50 | 100 |
| This work | 133 | 265 | 31.2 | 49 | 17 | 25 | 6.6 | 10.5 | 5 | 6.7 |
| Müller et al. (1993) | - | 270 | - | 49 | - | 19 | - | 8.9 | - | - |
| Möhlmann et al. (1979) | 224 | 355 | - | 68.3 | - | 27.3 | - | 10.2 | - | 4.16 |
| Beenakker et al. (1974) | - | - | 40.4 | 64.1 | - | - | - | - | - | - |

| Vroom et al. (1969) | 522 | 522 | 91.3 | 91.3 | 39.3 | 38.2 | 16.3 | 15.6 | - | - |

**Table 3** – Measured thresholds $E_{ap}$ and measured values previously reported in literature.

| λ (nm) | Transition | $E_{ap}$ (eV) | Reported (eV) | References |
|---|---|---|---|---|
| 283.5 | OH $A^2\Sigma^+ - X^2\Pi^+$ (1–0) | 9.4 ± 0.3 | | |
| 307.3 | OH $A^2\Sigma^+ - X^2\Pi^+$ (0–0) | 9.3 ± 0.3 | 9.0 ± 0.3 | Beenakker et al 1974 |
| | | | 9.2 ± 0.5 | Müller et al. 1993 (1) |
| | | | 9.0 ± 0.5 | Khodorkovskii et al. 2009 |
| 359.6 | OH$^+$ $A^2\Sigma^+ - X^2\Pi^+$ (0–0) | 23 ± 0.3 | | |
| 496.4 | H$_2$O$^+$ $\tilde{A}^2A_1 - \tilde{X}^2B_1$ | 15.8 ± 0.3 | 15 ± 0.5 | Müller et al. 1993 (1,2) |
| 503.3 | H$_2$O$^+$ $\tilde{A}^2A_1 - \tilde{X}^2B_1$ | 15.8 ± 0.3 | | |
| 519.8 | H$_2$O$^+$ $\tilde{A}^2A_1 - \tilde{X}^2B_1$ | 15.5 ± 0.3 | | |
| 656.9 | H$_\alpha$ (3 – 2) | 17.74 ± 0.3 | 18 ± 0.5 | Müller et al. 1993 (1) |
| | | | 18.5 ± 0.5 | Beenakker et al. 1974 |
| | | | 18.0 ± 0.5 | Böse & Sroka 1973 (3) |
| | | | 18.7 ± 0.4 | Ogawa et al. 1991 (4) |
| | | 23.6 ± 0.3 | 26.8 ± 1.5 | Beenakker et al. 1974 |
| | | | 25.5 ± 1.5 | Böse & Sroka 1973 (3) |
| | | | 25.5 ± 0.8 | Ogawa et al. 1991 (4) |
| | | | 31.3 ± 1.0 | Ogawa et al. 1991 (4) |
| | | | 38.9 ± 1.5 | Ogawa et al. 1991 (4) |

| | | | | | |
|---|---|---|---|---|---|
| 486.9 | $H_\beta$ (4 – 2) | | 18.46 ± 0.3 | 18 ± 0.5 | Müller et al. 1993 (1) |
| | | | | 18.6 ± 0.5 | Beenakker et al 1974 |
| | | | | 18.8 ± 0.5 | Böse & Sroka 1973 (3) |
| | | | | 18.7 ± 0.4 | Kurawaki et al. 1983 |
| | | | 25.5 ± 0.3 | 26 ± 0.5 | Müller et al. 1993 (1) |
| | | | | 26.8 ± 0.8 | Böse & Sroka 1973 (3) |
| | | | | 25.5 ± 0.8 | Kurawaki et al. 1983 |
| | | | - | 31.3 ± 1.0 | Kurawaki et al. 1983 |
| | | | - | 38.9 ± 1.5 | Kurawaki et al. 1983 |
| 434.6 | $H_\gamma$ (5 – 2) | | 18.6 ± 0.3 | 19.1 ± 0.5 | Beenakker et al 1974 |
| | | | | 18.9 ± 0.6 | Böse & Sroka 1973 (3) |
| | | | 25.5 ± 0.3 | 26.4 ± 0.8 | Böse & Sroka 1973(3) |
| 410.6 | $H_\delta$ (6 – 2) | | 19.45 ± 0.3 | 18.9 ± 0.5 | Böse & Sroka 1973 (3) |
| | | | 23.9 ± 0.3 | | Böse & Sroka 1973 (3) |
| | | | | 28.3 ± 1.3 | |
| 397.5 | $H_\epsilon$ (7 – 2) | | | 18.6 ± 0.5 | Böse & Sroka 1973 (3) |
| | | | | 29.0 ± 1.0 | Böse & Sroka 1973 (3) |

**Notes:** (1) Uncertainty based on half width of energy temp distribution; (2) Measured at 4614 A in the (0 -> 0, 2,0) band; (3) Based on the measured onset of the Lyman series; (4) Did not differentiate between and 4.

**Table 4** – Measured ($E_{ap}$) and calculated thermochemical ($E_{min}$) minimal appearance energies of emission features at wavelength $\lambda$.

| $\lambda$ (nm) | Transition | Products | $E_{min}$ (eV) | $E_{ap}$ (eV) |
|---|---|---|---|---|
| 260 – 360 | OH ($A^2\Sigma^+ - X^2\Pi^+$) | OH($A^2\Sigma^+$) + H | 9.24 | 9.4 ± 0.3 |
| | | OH($A^2\Sigma^+$) + H(2) | 19.45 | |
| 335 – 380 | OH$^+$ ($A^2\Sigma^+ - X^2\Pi^+$) | OH$^+$ ($A^2\Sigma^+$) + H | 21.66 | 23 ± 0.3 |
| | | OH$^+$ ($A^2\Sigma^+$) + H(2) | 31.87 | |
| 380 – 600 | H$_2$O$^+$ ($\tilde{A}^2A_1 - \tilde{X}^2B^1$) | H$_2$O$^+$ ($\tilde{A}^2A_1$) | 14.62 | 15.8 ± 0.3 |
| 656.3 | H$_\alpha$ (3 – 2) | OH, H(3) | 17.29 | 17.74 ± 0.3 |
| | | OH$^+$, H(3) | 20.80 | |
| | | OH($A^2\Sigma^+$) + H(3) | 21.34 | |
| | | O + H + H(3) | 21.74 | 23.6 ± 0.3 |
| | | OH$^+$($A^3\Pi$) + H(3) | 24.25 | |
| | | O($^1S^0$) + H + H(3) | 25.94 | |
| 486.1 | H$_\beta$ (4 – 2) | OH, H(4) | 17.94 | 18.46 ± 0.3 |
| | | OH$^+$, H(4) | 21.45 | |
| | | OH($A^2\Sigma^+$) + H(4) | 21.99 | |
| | | O + H + H(4) | 22.39 | |
| | | OH$^+$ ($A^3\Pi$) + H(4) | 24.90 | 25.5 ± 0.3 |
| | | O(1S0) + H + H(4) | 26.59 | |
| 434.0 | H$_\gamma$ (5 – 2) | OH, H(5) | 18.25 | 18.6 ± 0.3 |
| | | OH$^+$, H(5) | 21.76 | 21.6 ± 0.3 |
| | | OH(A$^2\Sigma^+$) + H(5) | 22.30 | |
| | | O + H + H(5) | 22.70 | |
| | | OH$^+$ (A$^3\Pi$) + H(5) | 25.21 | 25.5 |
| | | O(1S0) + H + H(5) | 26.90 | |
| 410.2 | H$_\delta$ (6 – 2) | OH, H(6) | 18.41 | 19.45 |
| | | OH$^+$, H(6) | 21.92 | |
| | | OH(A$^2\Sigma^+$) + H(6) | 22.46 | 23.9 |
| | | O + H + H(6) | 22.86 | |
| | | OH$^+$(A$^3\Pi$) + H(6) | 25.37 | |
| | | O(1S0) + H + H(6) | 27.06 | |
| 397.0 | H$_\epsilon$ (7 – 2) | OH, H(7) | 18.51 | |
| | | OH$^+$, H(7) | 22.02 | |
| | | OH(A$^2\Sigma^+$) + H(7) | 22.56 | |
| | | O + H + H(7) | 22.96 | |
| | | OH$^+$(A$^3\Pi$) + H(7) | 25.47 | |
| | | O(1S0) + H + H(7) | 27.16 | |

**Table 5** – Absolute emission cross sections $\sigma$ of OH (A $^2\Sigma^+$ - X $^2\Pi$) for 50 eV electron impact on water. 'Unresolved' implies the total emission cross sections of bands within the emission features at 307 –330 nm (top three rows) and 280 – 295 nm (next two rows). The uncertainties of the shown values are approximately 20% (see sec. 2.4 for details).

| $v' - v''$ | $\sigma$ | Unresolved (model) | Unresolved (experiment) at 50 eV | Muller (1993) (unresolved) at 100 eV |
|---|---|---|---|---|
| | (cm$^2$) | (cm$^2$) | (cm$^2$) | (cm$^2$) |
| (0 – 0) | 1.14 x 10$^{-18}$ | | | |
| (1 – 1) | 2.30 x 10$^{-19}$ | 1.38 x 10$^{-18}$ | 1.39 x 10$^{-18}$ | 2.68 x 10$^{-18}$ |
| (2 – 2) | 8.56 x 10$^{-21}$ | | | |
| (1 – 0) | 1.39 x 10$^{-19}$ | 1.52 x 10$^{-19}$ | 1.56 x 10$^{-19}$ | 3.4 x 10$^{-19}$ |
| (2 – 1) | 1.34 x 10$^{-20}$ | | | |
| (2 – 0) | 1.83 x 10$^{-21}$ | - | 4.64 x 10$^{-21}$ | - |

**Table 6** - Absolute emission cross sections $\sigma$ in cm$^2$ of OH$^+$ (A $^3\Pi$ – X $^3\Sigma^-$) and H$_2$O$^+$ (A$^2$A$_1$ – X$^2$B$_1$) for electron impact on water at incident energies of 50 eV. Due to weak intensities and band overlaps we estimate the uncertainty of the shown values to be approximately 30%.

| Transition | Integration boundaries (nm) | Experimental $\sigma$ (cm$^2$) |
|---|---|---|
| OH$^+$ (A-X)(0-0) | 356.7 – 367.0 | 2.04 x 10$^{-20}$ |
| OH$^+$ (A-X)(1-0) | 333.3 – 340.8 | 1.41 x 10$^{-20}$ |
| OH$^+$ (A-X)(1-1) | 367.2 – 378.0 | 1.20 x 10$^{-20}$ |
| OH$^+$ (A-X)(2-1) | 346.6 – 356.0 | 1.75 x 10$^{-20}$ |
| H$_2$O$^+$(A-X)(0,12,0)-(0,0,0) | 495.8 – 496.6 | 2.76 x 10$^{-20}$ |
| H$_2$O$^+$(A-X)(0,13,0)-(0,1,0) | 502.8 – 503.8 | 2.63 x 10$^{-20}$ |
| H$_2$O$^+$(A-X)(0,11,0)-(0,0,0) | 519.3 – 520.3 | 2.46 x 10$^{-20}$ |

**Table 7** – Relative line strengths of the Hydrogen Balmer series with respect to $H_\alpha$ for electron impact on water vapor at energies of 50 and 100 eV (cf. Table 2), compared to resonant fluorescence by hydrogen atoms.

| Transition | Relative strength compared to $H_\alpha$ | | |
|---|---|---|---|
| | Electron Impact at 50 eV | Electron Impact at 100 eV | Fluorescence |
| $H_\alpha$ (3-2) | 1.00 | 1.00 | 1.00 |
| $H_\beta$ (4-2) | 0.23 | 0.19 | 0.23 |
| $H_\gamma$ (5-2) | 0.13 | 0.09 | 0.01 |
| $H_\delta$ (6-2) | 0.05 | 0.04 | 0.02 |
| $H_\varepsilon$ (7-2) | 0.04 | 0.03 | 0.01 |

# Appendix

**Table A1** – Parameters of the polynomial fits of the determined excitation-emission cross section curves. The curves were fitted by polynomial function $y = A_0 + A_1.x + A_2.x^2 + … + A_9.x^9$. The fits are valid only in the electron energy range given.

|  | H(3-2) | H(4-2) | H(4-2) | H(5-2) |
|---|---|---|---|---|
|  | Range: 17.74 – 100 eV | Range: 18.46 – 25 eV | Range: 25– 100 eV | Range: 18.6 – 100 eV |
| $A_0$ | 1.166448197771539e-21 | 7.639860613710967e-16 | 4.943728915208466e-17 | -1.3340396166601772e-18 |
| $A_1$ | 9.303043241829306e-20 | -4.339906653651744e-16 | -8.284530518495484e-18 | 3.142413051829726e-19 |
| $A_2$ | -1.047964582351875e-20 | 1.0844583626545231e-16 | 5.932055298901659e-19 | -3.087841795930737e-20 |
| $A_3$ | 2.828752787633401e-22 | -1.5639628844033396e-17 | -2.390354805210368e-20 | 1.6552067983794612e-21 |
| $A_4$ | 1.608911911675589e-23 | 1.4341280890695941e-18 | 6.003860671654366e-22 | -5.323693565345372e-23 |
| $A_5$ | -1.087476686916897e-24 | -8.669350340974055e-20 | -9.783989360655368e-24 | 1.0748791030573578e-24 |
| $A_6$ | 2.6022714570262697e-26 | 3.454204282746568e-21 | 1.0379278535794464e-25 | -1.3740836557017882e-26 |
| $A_7$ | -3.120952424581023e-28 | -8.746607094179829e-23 | -6.931564067835833e-28 | 1.0801017593637847e-28 |
| $A_8$ | 1.8805374063667342e-30 | 1.2772175870694166e-24 | 2.6503850787057475e-30 | -4.76448663594286e-31 |
| $A_9$ | -4.5367494258922786e-33 | -8.195318726717443e-27 | -4.4286991283227496e-33 | 9.028240042967173e-34 |

**Table A2** – Parameters of the polynomial fits of the determined excitation-emission cross section curves. The curves were fitted by polynomial function $y = A_0 + A_1 \cdot x + A_2 \cdot x^2 + \ldots + A_9 \cdot x^9$. The fits are valid only in the electron energy range given.

|  | H(6-2) |  | H(7-2) | OH(A-X)(0-0) | OH(A-X)(1-0) |
|---|---|---|---|---|---|
|  | Range: 19.45 – 25 eV | Range: 25 – 100 eV | Range: 20 – 100 eV | Range: 9.3 – 100 eV | Range: 9.4 – 100 eV |
| $A_0$ | 4.6050832232610517e-20 | 2.7887405837723353e-18 | 2.5224214271660094e-22 | -1.6708105463999309e-18 | -1.6068108298970508e-18 |
| $A_1$ | -9.374373347463644e-21 | -4.625627979181022e-19 | -9.344411261004822e-22 | -5.003481239748052e-19 | 3.591444305340944e-19 |
| $A_2$ | -4.7778670332226695e-21 | 3.290421775742448e-20 | 2.0011358601925912e-22 | 1.501604107447954e-19 | -3.02044645641766e-20 |
| $A_3$ | 2.3619183204472313e-21 | -1.3201234056221827e-21 | -1.6159383332511198e-23 | -1.1353168952795955e-20 | 1.4454391127131222e-21 |
| $A_4$ | -4.447622886730001e-22 | 3.3050664777090135e-23 | 7.094824362753076e-25 | 4.265395602292418e-22 | -4.3647020503110567e-23 |
| $A_5$ | 4.610214959628704e-23 | -5.369524359823974e-25 | -1.8453223153121365e-26 | -9.315104529482383e-24 | 8.588802513522081e-25 |
| $A_6$ | -2.8512295407360197e-24 | 5.673896248847104e-27 | 2.9162264157020105e-28 | 1.2388681669547896e-25 | -1.09700249249277756e-26 |
| $A_7$ | 1.0487461432786806e-25 | -3.7674680609698995e-29 | -2.7467021627222484e-30 | -9.900092258118275e-28 | 8.753050817957322e-29 |
| $A_8$ | -2.1181961623891444e-27 | 1.4285969698096938e-31 | 1.416522775284146e-32 | 4.3712669183598735e-30 | -3.957593265862472e-31 |
| $A_9$ | 1.8088489514284093e-29 | -2.359940585253714e-34 | -3.0761767267104824e-35 | -8.196452209932972e-33 | 7.729341580431711e-34 |

**Table A3** – Parameters of the polynomial fits of the determined relative excitation-emission curves. The curves were fitted by polynomial function $y = A_0 + A_1 \cdot x + A_2 \cdot x^2 + \ldots + A_9 \cdot x^9$. The fits are valid only in the shown electron energy range.

|  | $OH^+(A-X)(0-0)$ | $H_2O^+$ (A-X) 496.4nm | $H_2O^+$ (A-X) 503.3nm | $H_2O^+$ (A-X) 519.8nm |
|---|---|---|---|---|
|  | Range: 23 – 100 eV | Range: 15.8 – 100 eV | Range: 15.8 – 100 eV | Range: 15.5 - 100eV |
| $A_0$ | -4.879653950408096e-21 | 8.858789956334654e-20 | 8.393390382947733e-20 | -1.1149998139715028e-19 |
| $A_1$ | 2.659247657749363e-21 | -1.9908158448335207e-20 | -2.1597167385367772e-20 | 2.2894976608348353e-20 |
| $A_2$ | -1.7627911756207753e-22 | 1.7733554480488418e-21 | 2.1710745284722074e-21 | -2.0344880667853087e-21 |
| $A_3$ | -1.6520163112711283e-24 | -8.131201536101095e-23 | -1.1261321281605362e-22 | 1.0594033792359092e-22 |
| $A_4$ | 6.342836468753547e-25 | 2.172870150769907e-24 | 3.4685543164993574e-24 | -3.436446082576604e-24 |
| $A_5$ | -2.769228380473545e-26 | -3.5155158139396874e-26 | -6.674381095138968e-26 | 7.137233719051186e-26 |
| $A_6$ | 5.696303051428173e-28 | 3.403293871656412e-28 | 8.089104069469202e-28 | -9.480450797570219e-28 |
| $A_7$ | -6.2909781392886665e-30 | -1.8341384208676137e-30 | -5.9900506073436436e-30 | 7.778923273105029e-30 |
| $A_8$ | 3.6037202657997516e-32 | 4.456416074602091e-33 | 2.471321253124862e-32 | -3.5874248127206663e-32 |
| $A_9$ | -8.423875581896907e-35 | -1.7179486857965017e-36 | -4.3454776734175923e-35 | 7.106174063364234e-35 |